 \documentclass[final,3p,times,sort&compress]{elsarticle}

\usepackage{multirow,setspace,times,amssymb,amsmath,graphicx,color,rotating,subfigure,url}
\usepackage{lineno}
\usepackage{natbib}
\usepackage{booktabs}
\usepackage{hyperref}

\journal{Physica A}

\begin{document}

\begin{frontmatter}


\title{Short-term Market Reaction after Trading Halts in Chinese Stock Market}
\author[CME,CCSC]{Hai-Chuan Xu}
\author[CME,CCSC]{Wei Zhang}
\author[CME,CCSC]{Yi-Fang Liu \corref{cor}}
\cortext[cor]{Corresponding author. Address: Rm.713 Beiyang Science Building, College of Management and Economics,
              Tianjin University, Tianjin 300072, China, Phone: +86 22 27891308.}
\ead{yifang.0731@gmail.com} %

\address[CME]{College of Management and Economics, Tianjin University, Tianjin, 300072, China}
\address[CCSC]{China Center for Social Computing and Analytics, Tianjin University, Tianjin, 300072, China}

\begin{abstract}
  In this paper, we study the dynamics of absolute return, trading volume and bid-ask spread after the trading halts using high-frequency data from the Shanghai Stock Exchange. We deal with all three types of trading halts, namely intraday halts, one-day halts and inter-day halts, of 203 stocks in Shanghai Stock Exchange from August 2009 to August 2011. We find that absolute return, trading volume, and in case of bid-ask spread around intraday halts share the same pattern with a sharp peak and a power law relaxation after that. While for different types of trading halts, the peaks\textquoteright{} height and the relaxation exponents are different. From the perspective of halt reasons or halt durations, the relaxation exponents of absolute return after inter-day halts are larger than those after intraday halts and one-day halts, which implies that inter-day halts are most effective. From the perspective of price trends, the relaxation exponents of excess absolute return and excess volume for positive events are larger than those for negative events in case of intraday halts and one-day halts, implying that positive events are more effective than negative events for intraday halts and one-day halts. In contrast, negative events are more effective than positive events for inter-day halts.
\end{abstract}


\end{frontmatter}

\section{Introduction}
\label{S1:Introduction}

Financial markets are complex systems characterized by emerging extreme events \cite{Sornette-2009-PUP}. It is a meaningful thing to answer the question how financial market dynamics are affected when the system undergoes an extreme event, such as financial crash, interest rate shock, large price change or trading halt. In recent years, statistical physics are applied to understand these financial markets dynamics, which are discovered sharing the feature of power law distributions. Early works were done by Lillo and Mantegna, who focused on relaxation dynamics of aftershocks after a crash. They researched the 1-min logarithm changes of Standard and Poor\textquoteright s 500 index during 100 trading days after the Black Monday and discovered the decaying patterns in the rate of aftershocks larger than some threshold, as Omori\textquoteright s law after earthquakes \cite{Lillo-Mantegna-2003-PRE,Lillo-Mantegna-2004-PA}. Weber et al. further found after crash period is characterized by the Omori process on all scales \cite{Weber-Wang-Vodenska-Havlin-Stanley-2007-PRE}. They studied the 1-min return series of three different data sets: S$\&$P 500 index in the 15000 trading minutes after Black Monday on 19 October 1987; quotes of 100 most frequently traded stocks at NYSE, NASDAQ and AMEX in two months after the crash on 27 October 1997; return series of General Electric stock in three months after 11 September 2001. In addition, the relaxation dynamics of aftershocks after large volatilities rather than large crashes were also investigated and similarly decay as a power law \cite{Mu-Zhou-2008-PA,Petersen-Wang-Havlin-Stanley-2010-PRE-036114}. Furthermore, the aftershocks dynamics after U.S. Federal Open Market Committee meetings that will announce interest rate change was described as an analog of the Omori law \cite{Petersen-Wang-Havlin-Stanley-2010-PRE-066121}. It is an example of aftershock dynamics that is clearly due to an external event. Apart from the dynamic of occurrence rate of aftershocks, another related topic is the relaxation dynamics of some financial measures after large price change / large bid-ask spread change. Zawadowski et al. examined high frequency data from NYSE and NASDAQ to conclude that volatility, volume and in case of the NYSE bid-ask spread increase sharply at the large intraday price change and decay according to a power law \cite{Zawadowski-Kertesz-Andor-2004-PA,Zawadowski-Andor-Kertesz-2006-QF}. Ponzi et al. studied the dynamics of the bid-ask spread and the mid-price after a sudden variation of spread, and then found that the spread decays as a power law to its normal value \cite{Ponzi-Lillo-Mantegna-2009-PRE}. Moreover, the order flow measures, such as the volume of different types of orders and the volume imbalance, were also discovered to increase before extreme events and decay slowly as a power law \cite{Mu-Zhou-Chen-Kertesz-2010-NJP,Toth-Kertesz-Farmer-2009-EPJB}. These dynamics can also be viewed as a type of switching phenomena in financial markets \cite{Stanley-Buldyrev-Franzese-2010-PA,Preis-Stanley-2010-JSP,Preis-Schneider-Stanley-2011-PNAS}. These researches illustrate the scale-free behavior of trading volume both before and after the end of a trend \cite{Stanley-Buldyrev-Franzese-2010-PA,Preis-Schneider-Stanley-2011-PNAS}. A significant sudden jump of the volatility and then a power law decay can also be found for both microtrends in the German DAX future time series and macrotrends in the daily closing price time series of all S$\&$P 500 stocks \cite{Preis-Stanley-2010-JSP}. However, the discovery of Jiang et al. is a bit of different, that is, the volatility dynamics both before and after large fluctuations are symmetric in time scales of minutes, while asymmetric in daily time scales \cite{Jiang-Chen-Zheng-2013-PA}. These analyses reveal that the asymmetry is mainly induced by exogenous shocks, whose precursory and relaxation dynamics in social systems have been studied by \cite{Sornette-Helmstetter-2003-PA,Sornette-Deschatres-Gilbert-Ageon-2004-PRL,Roehner-Sornette-Andersen-2004-IJMPC,Deschatres-Sornette-2005-PRE,Crane-Sornette-2008-PNAS}.

So far researchers have studied the dynamics around financial crash, large price changes or interest rate changes with the technique of statistical analysis in high-resolution data. This paper will study another kind of extreme events commonly occurred in stock markets, namely trading halts, from this point of view. Trading halt is one of microstructure mechanisms in equity market designed to temporarily stop trading during the period of extremely price movement or of the announcement of significant events. Some financial literatures \cite{Lee-Ready-Seguin-1994-JF,Subrahmanyam-1994-JF,Corwin-Lipson-2000-JF,Ackert-Church-Jayaraman-2001-JFM,Christie-Corwin-Harris-2002-JF,Edelen-Gervais-2003-RFS,Engelen-Kabir-2006-JBFA,Hauser-Kedar-Pilo-Shurki-2006-JFSR,Madura-Richie-Tucker-2006-JFSR,Jiang-McInish-Upson-2009-JFM,Chakrabarty-Corwin-Panayides-2011-JFI} have studied the effects of trading halts, such as price discovery, liquidity and volatility, from the perspective of information dissemination and transaction cost. While no unified pattern has been discovered and few studies focus on comparing the dynamics around different types of trading halts. Therefore, we attempt to find the unified behavior of different financial measures and to compare the relaxation dynamics around different trading halts.

In this work, we investigate the relaxation dynamics of several financial measures around different types of trading halts in Chinese stock market. This paper is organized as follows. The next section provides data description. Section 3 characterizes the dynamics of cumulative return. Section 4 investigates the relaxation dynamics of absolute return, trading volume and bid-ask spread after trading halts. Section 5 studies the power law behavior after trading halts. Conclusions are provided in Section 6.

\section{Data sets}
\label{S1:Data}

We analyze 1-min high-frequency trading data of 203 stocks traded on Shanghai Stock Exchange (SSE) between August 2009 and August 2011. These stocks are selected based on scale and liquidity and cover a variety of industry sectors. The tickers of these 203 stocks are listed in Appendix. In this paper, we only consider the trading occurring in the continuous double auction (9:30 to 11:30 AM and 1:00 to 3:00 PM every day). In Shanghai Stock Exchange (SSE), trading halts can be generally divided into 3 types depending on suspension reasons: halts due to abnormal price fluctuations, which last for 1 hour and here we call intraday halts; halts due to shareholders\textquoteright{} meeting, which last for one day and we call one-day halts; halts due to announcement of significant events, which last for more than one day and we call inter-day halts. A detailed survey about circuit breakers in financial markets pointed out that the firm-specific trading halts could generally be classified into 2 types: order imbalance halts and news related halts\cite{kim2004makes}. In this paper, the intraday halts are more or less similar with the order imbalance halts considering the fact that abnormal price fluctuations should be associated with order imbalance. However, it is not exact to equal these two types of halts because the order imbalance halts are designed to protect specialists or market makers, while there are no specialists and market makers in Shanghai Stock Exchange. According to the definitions above, inter-day halts can be viewed as news related halts and one-day halts are routine suspensions. With the given sample stocks and periods, we find 1341 trading halts. After removing the cases like successive halts, ST stocks halts, halts lasting for more than one month and data error or missing, there are 640 eligible trading halts left. Table~\ref{Tab:Type} gives the statistics of different types of trading halts, in which stock price shows an upward trend in the length of 240 minutes before the occurrence of halt is defined as a positive event, conversely is defined as a negative event\footnote{We chose $\Delta t =\rm{60,120,180,240}$ minutes as the length of time windows to calculate the price trend and do the following analysis, respectively. We found the reaction patterns and the comparative analysis are robust. In this paper, we only display the results for $\Delta t =\rm{240}$ minutes.}. We can find that the amount of one-day halts is much larger than that of intraday halts and inter-day halts. Meanwhile, the amount of negative events is more than that of positive events, which indicates that trading halts mostly occurs at the time when the company exist potential negative news.

\begin{table}[ht]
  \caption{\label{Tab:Type} Amounts of different types of trading halts.}
  \centering
  \begin{tabular}{c c c c}
  \toprule
   & & Events & \\ \cline{2-4}
  Types of trading halt & Total & Positive & Negative \\ \midrule
  Intraday & 24 & 14 & 10 \\
  One-day & 573 & 84 & 489 \\
  Inter-day & 43 & 28 & 15 \\
  Total & 640 & 126 & 514 \\
  \bottomrule
  \end{tabular}
\end{table}

\section{Dynamics of average cumulative return}
\label{S1:DynamicsCumulRet}

\begin{figure}[!htb]
  \centering
  \includegraphics[width=0.45\textwidth]{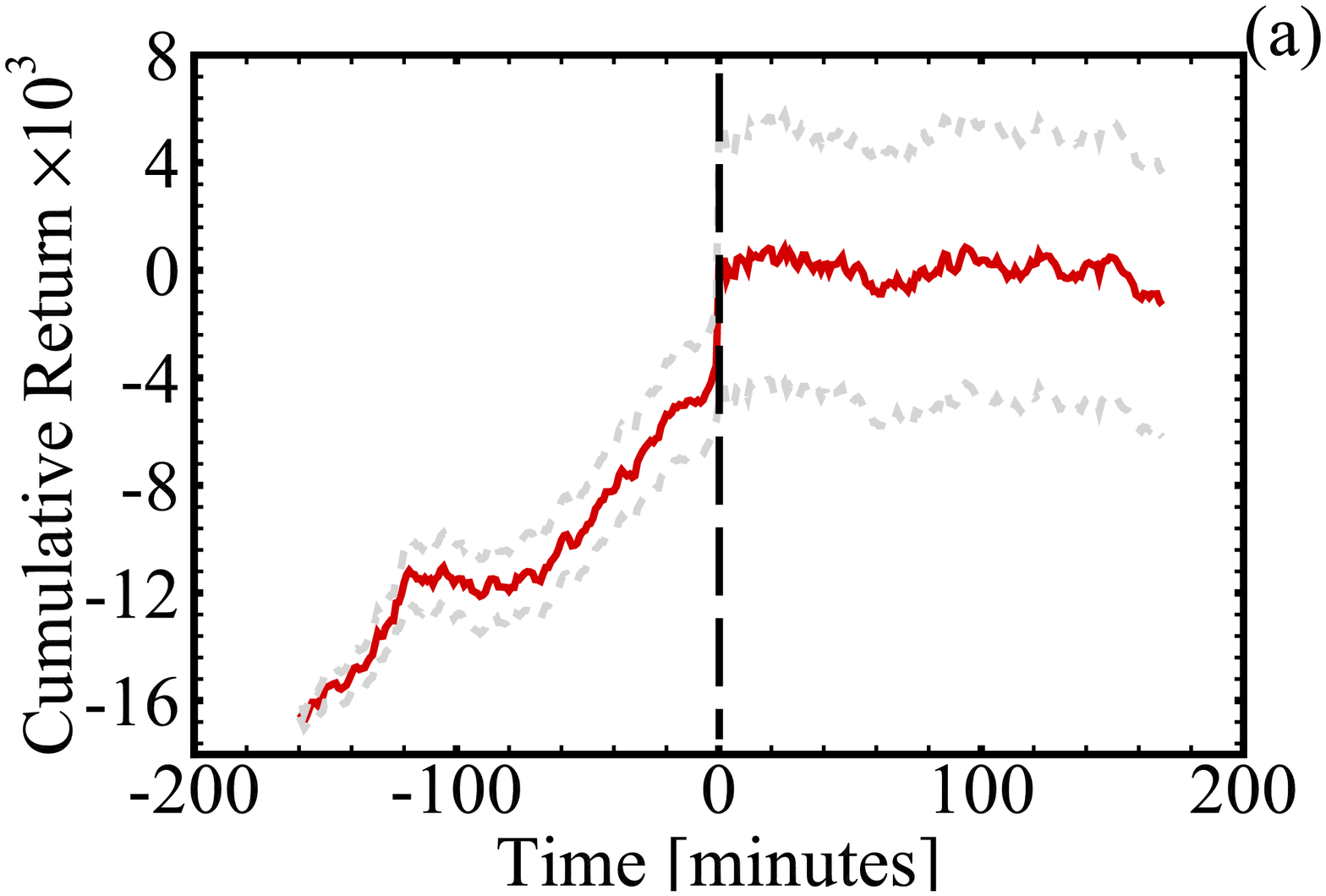}
  \includegraphics[width=0.45\textwidth]{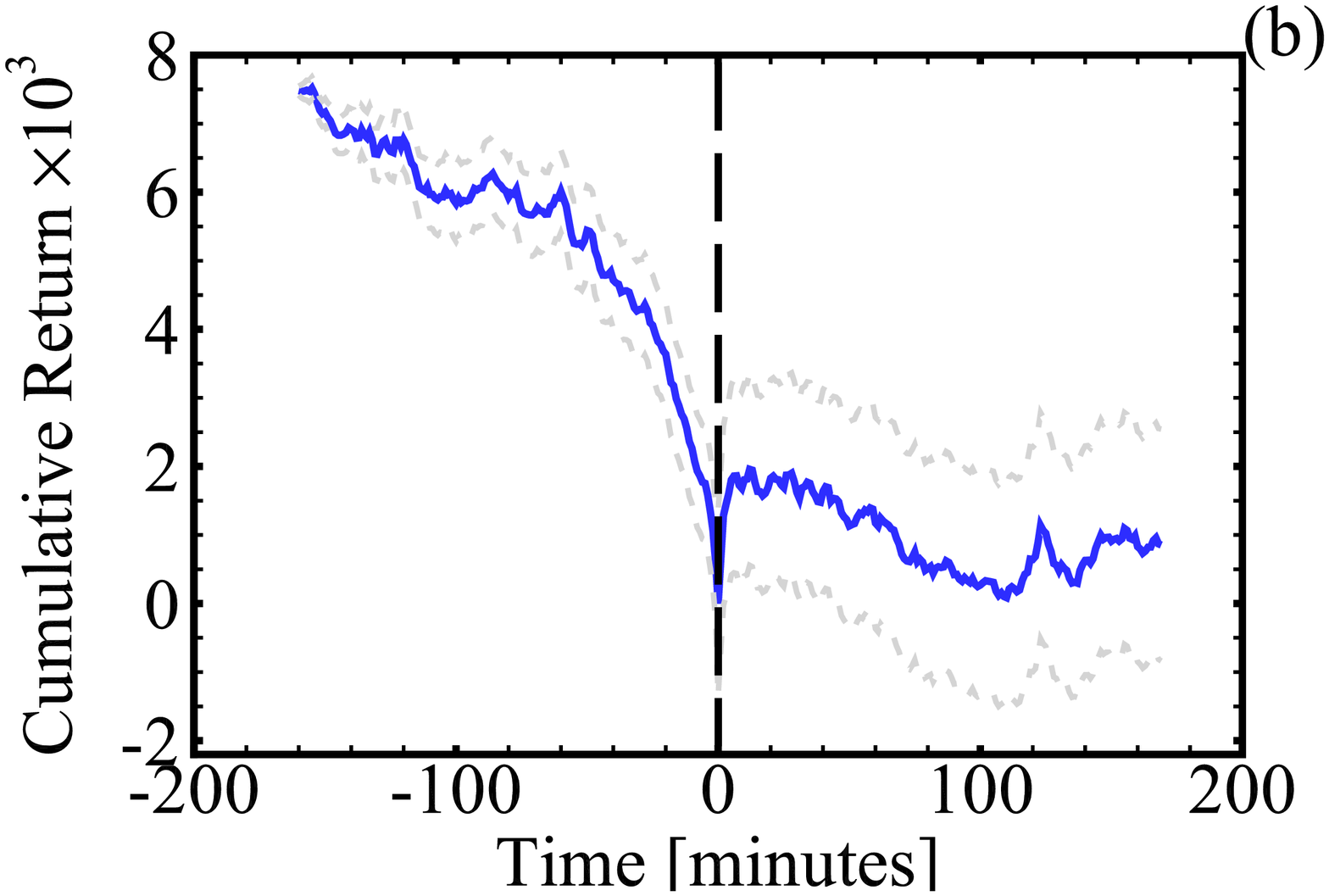}
  \caption{\label{Fig:CumRet:All} Average cumulative return averaged over 126 positive events with price increase (a) and 514 negative events with price decrease (b), respectively. Minute 0 corresponds to the time when the trading resumed. The gray lines demonstrate standard errors associated with each data point.}
\end{figure}

We will first analyze the dynamics of average cumulative return which reflect the effect of price discovery after trading halts. Figure~\ref{Fig:CumRet:All} gives the average cumulative return for 126 positive events and 514 negative events. The individual return is defined as
\begin{equation}
 r_t = \ln p_t - \ln p_{t-1},
 \label{Eq:LogReturn}
\end{equation}
where $p_t$ is the last transaction price within the $t$th minute. The average cumulative return is defined as
\begin{equation}
 R_t = \frac{1}{\|M\|} \sum_{m\in M} \sum_{i=-160}^t r_m (i),     t=\rm{-160,\ldots,160} ,
 \label{Eq:AvCumuReturn}
\end{equation}
where $\|M\|$ is the amount of trading halts in events group $M$. All events have equal weight in the averaging procedure. After averaging, the two curves are shifted vertically so that the cumulative returns at $t=\rm{0}$ are zero. We can find before occurrence of trading halts, the two curves go upwards and downwards respectively. After trading halts, in case of positive events, the cumulative return, thus the logarithmic price stops rising and maintains at a fairly stable value quickly. Similarly, in case of negative events, trading halts not only prevent the falling of cumulative return, but also reverse the cumulative return in the short time immediately after the halts. In other words, there are 326 individual stocks (63.4\%) within 1 minute and 460 individual stocks (89.5\%) within 2 minutes demonstrate reversal for the case of negative events. This indicates that trading halts play a certain role in price discovery.

\begin{figure}[!htb]
  \centering
  \includegraphics[width=0.32\textwidth]{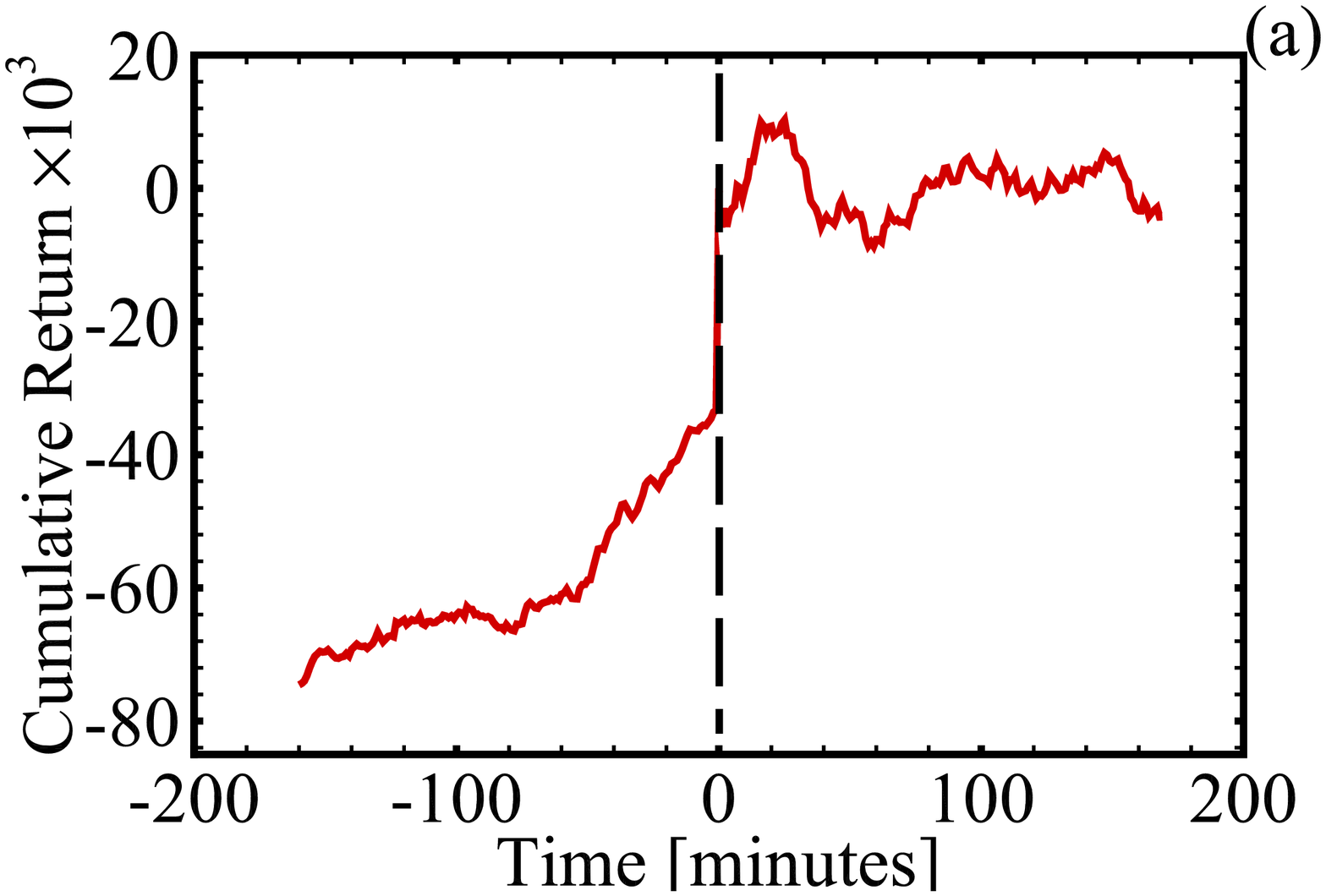}
  \includegraphics[width=0.32\textwidth]{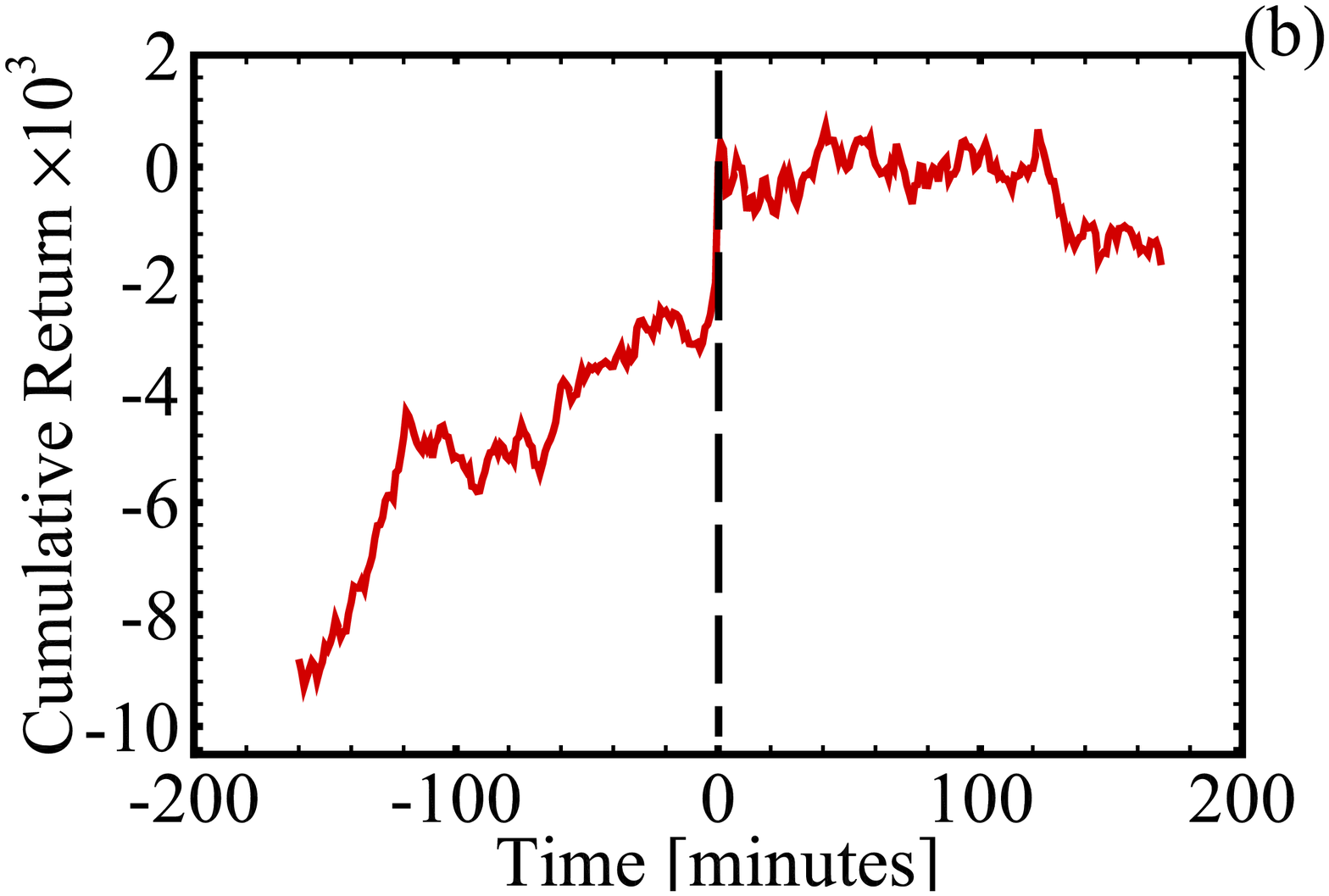}
  \includegraphics[width=0.32\textwidth]{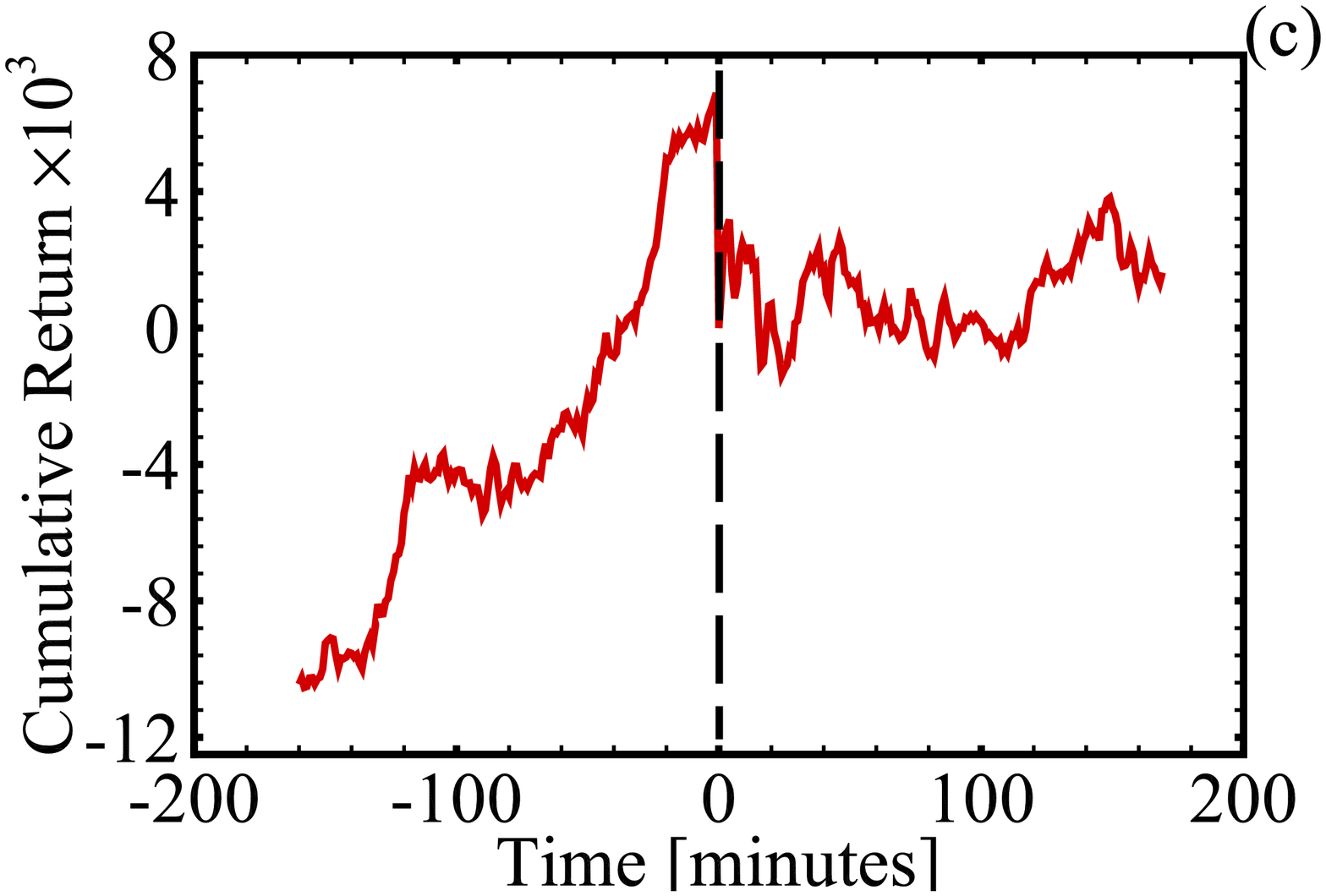}
  \includegraphics[width=0.32\textwidth]{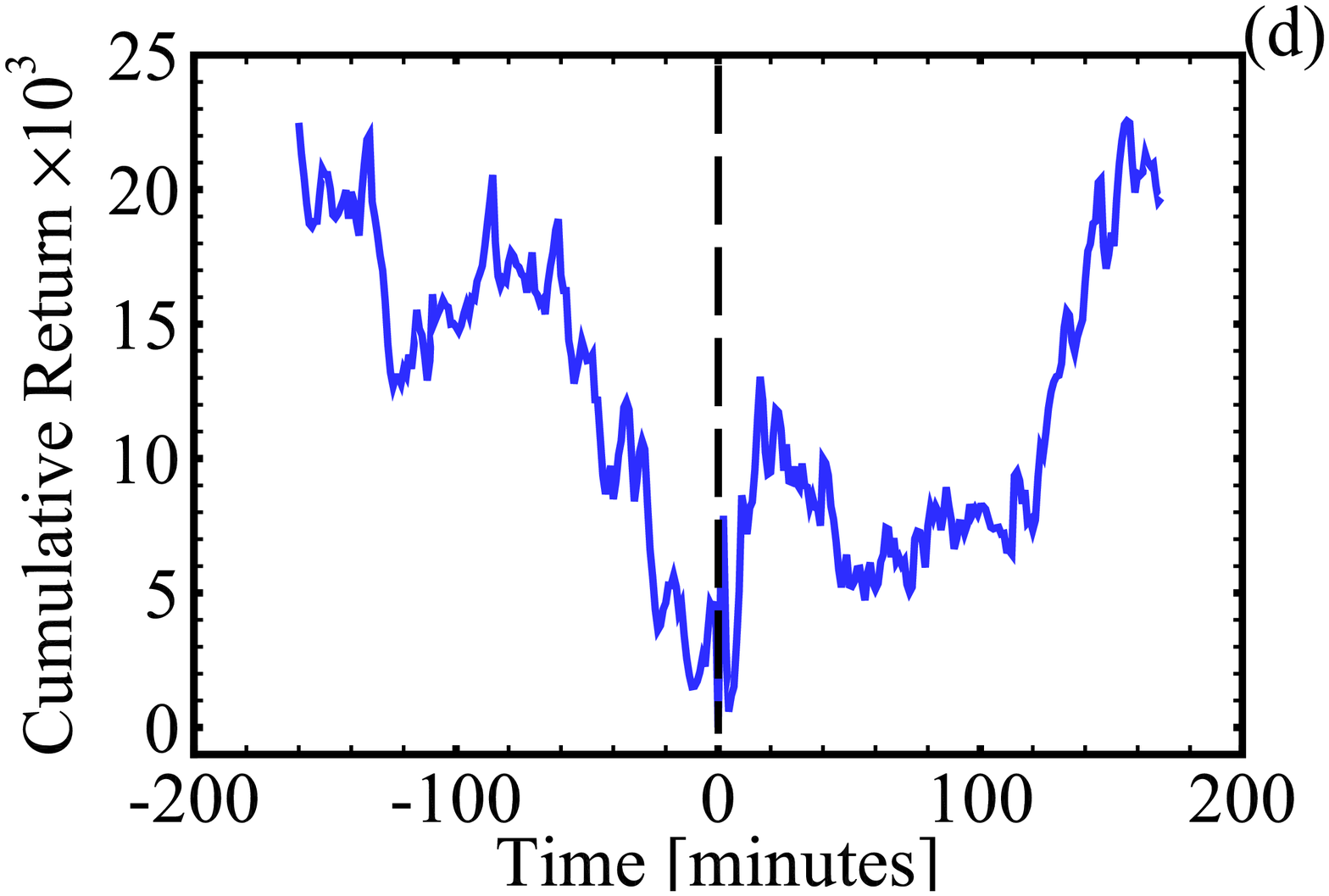}
  \includegraphics[width=0.32\textwidth]{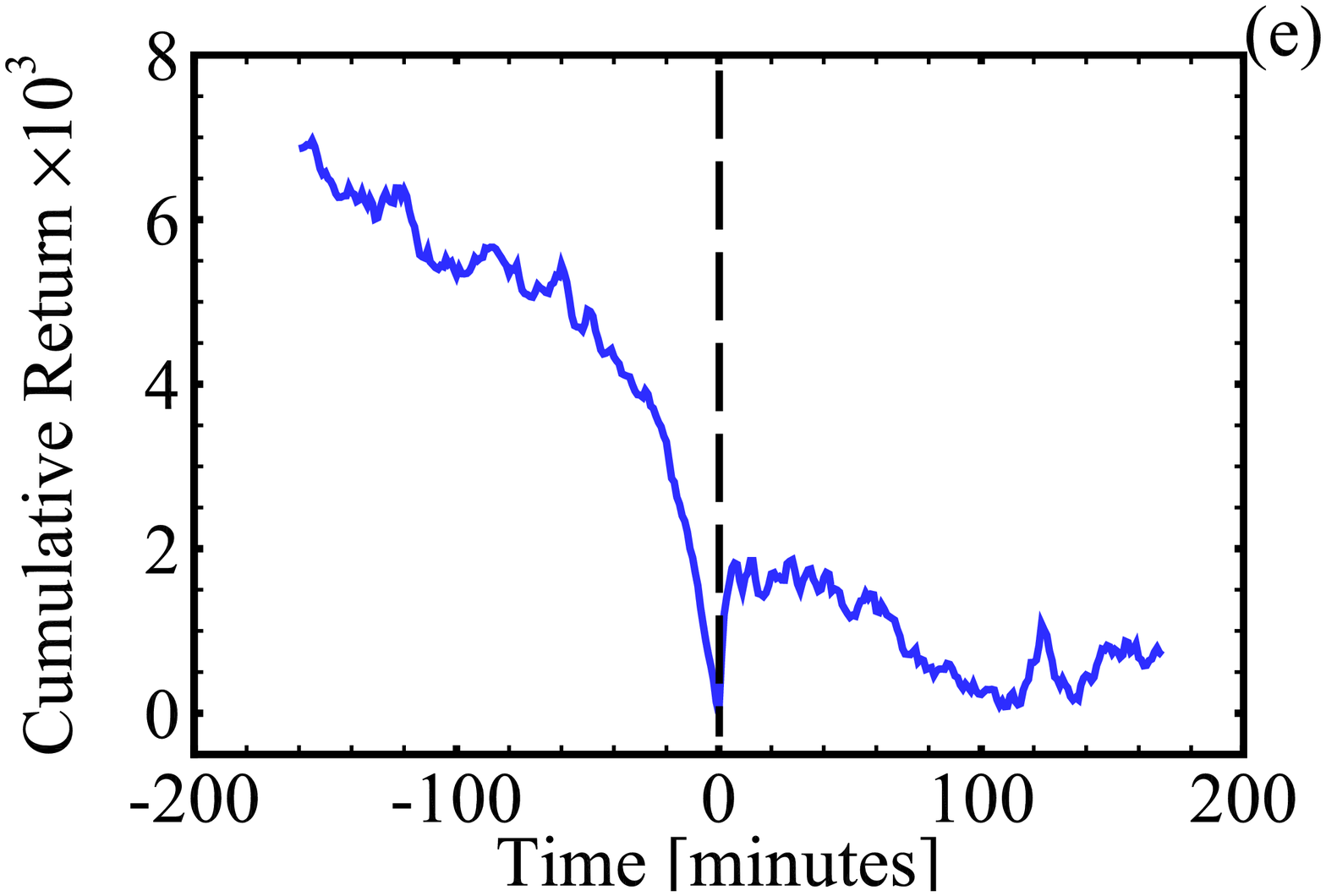}
  \includegraphics[width=0.32\textwidth]{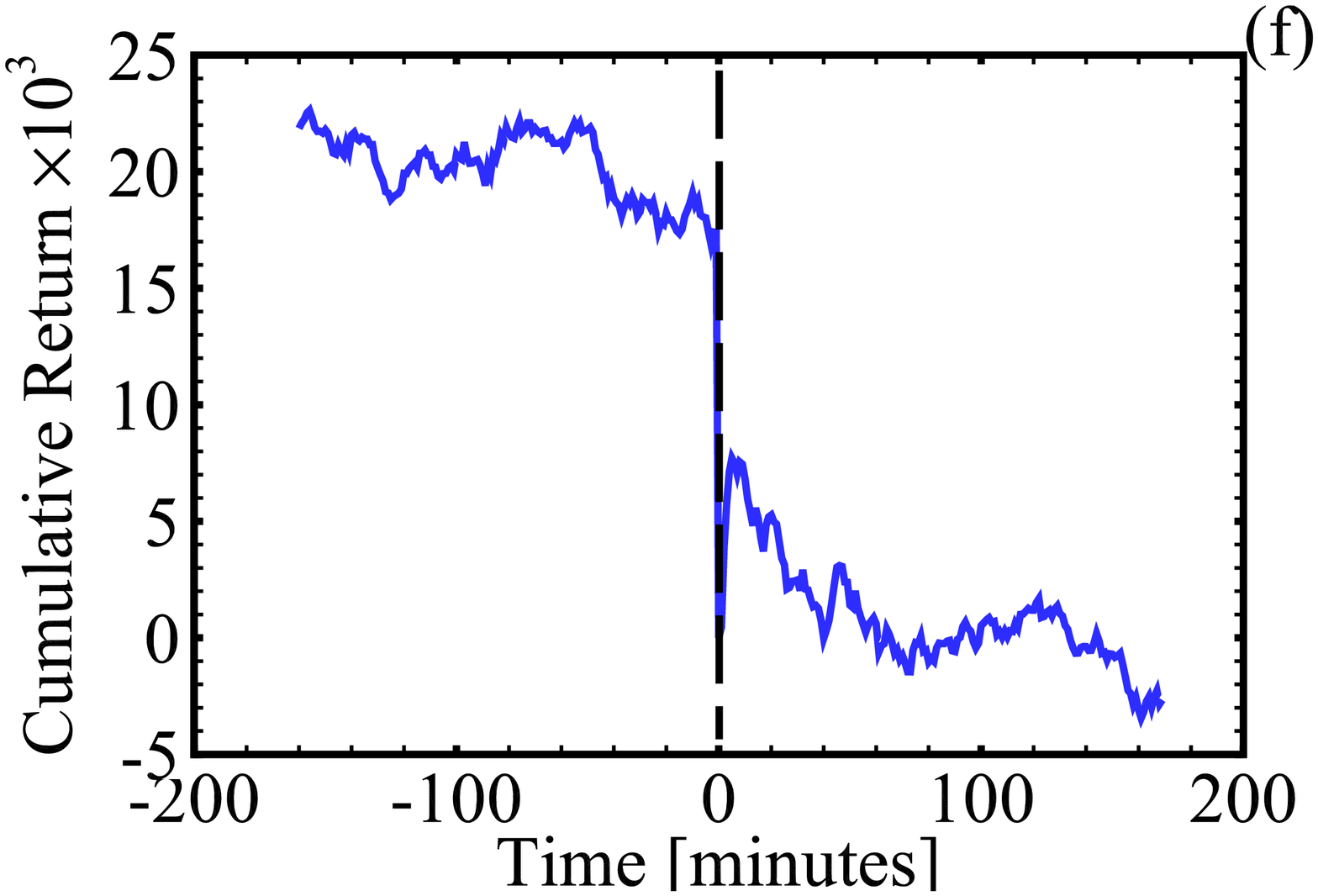}
  \caption{\label{Fig:CumRet:Each} Dynamics of average cumulative return around three types of trading halts which are also divided into positive and negative events: (a) intraday halts around 14 positive events, $s=0.0042$, (b) one-day halts around 84 positive events, $s=0.0006$, (c) inter-day halts around 28 positive events, $s=0.0012$, (d) intraday halts around 10 negative events, $s=0.0052$, (e) one-day halts around 489 negative events, $s=0.0005$, and (f) inter-day halts around 15 negative events, $s=0.0023$.}
\end{figure}

Furthermore, to compare the effect of price discovery, we investigate dynamics of cumulative return when different types of trading halts are imposed. Figure~\ref{Fig:CumRet:Each}(a)-~\ref{Fig:CumRet:Each}(c) show that the cumulative returns around all 3 types of positive halts display the similar pattern as in Figure~\ref{Fig:CumRet:All}(a). However, although all types of positive halts prevent the sustained rise of cumulative return, the stabilities in after-halt period are different. For each plot, we calculated the standard deviation $s$ for average cumulative returns after time 0 and found that the average cumulative returns after intraday halts are most volatile, while those after one-day halts are most stable. This difference may relate with the complexity of the information contained in the announcement. The intraday halts triggered by abnormal price fluctuations disseminate the most ambiguous information, while the inter-day halts triggered by announcing significant events imply relatively clearer massage and the one-day halts triggered by shareholders\textquoteright{} meeting distribute almost no uncertain information. For negative halts (Figure~\ref{Fig:CumRet:Each}(d)-~\ref{Fig:CumRet:Each}(f)), although the stabilities in after-halt period are consistent with the positive halts, which can also be explained by the complexity of the information, the shapes of these curves are a bit of different, especially for the intraday negative halts. The cumulative return after intraday halts (Figure~\ref{Fig:CumRet:Each}(d)) displays a reversion, indicating an overreaction is occurred, which reflects investors\textquoteright{} panic behavior when facing abnormal negative price fluctuations.

\section{Dynamics of three financial measures}
\label{S1:DynamicsThreeMeasures}

Now we will give some empirical results around different types of trading halts. Cumulative return itself is not the only important measure which gives us information on market behavior after the halt, so we will analyze three different measures, namely absolute return, trading volume and bid-ask spread. The absolute return is the absolute value of logarithmic return, which reflects the fluctuation of time series; trading volume is the quantity of all transactions within 1 minute, which reflects traders' aggressiveness or prudential investment; bid-ask spread is the difference between best ask price and best bid price at the end of each minute, which reflects transaction costs and market liquidity. Before aggregating all the trading halts in each category, we remove the intraday pattern from all investigated measures as \cite{Mu-Zhou-Chen-Kertesz-2010-NJP}. For each measure $Z(d,t')$, the intraday pattern $I(t')$ is determined as the average of measures at the same intraday time backtracking 40 days prior to the trading halt, where $d$ identifies trading days and $t'\rm{=1,2,3,\ldots,240}$ is the intraday time. The relative value removed intraday pattern from $Z(d,t')$ is
\begin{equation}
 z(d,t') = Z(d,t') / I(t'),
 \label{Eq:RemovePatten}
\end{equation}
for each trading day $d$. For each trading halt $h$ , the evolutionary trajectory $\{z_h (t):t\rm{=-80,-79,\ldots,-1,0,1,\ldots,159,160}\}$ is extracted from $z(d,t' )$, which contains 80 min before halt $h$ and 160 min after resumption of trading. Then we obtain the average of measure $Z$ for a group of trading halts $H$:
\begin{equation}
 z_H(t) = \frac{1}{\|H\|} \sum_{h\in H} z_h (t),     t=\rm{-80,\ldots,160} ,
 \label{Eq:AverageMeasure}
\end{equation}
where $\|H\|$ is the amount of trading halts in group $H$. Note that $t=\rm{0}$ corresponds to the time when the trading resumed.

\begin{figure}[htb]
  \centering
  \includegraphics[width=0.32\textwidth]{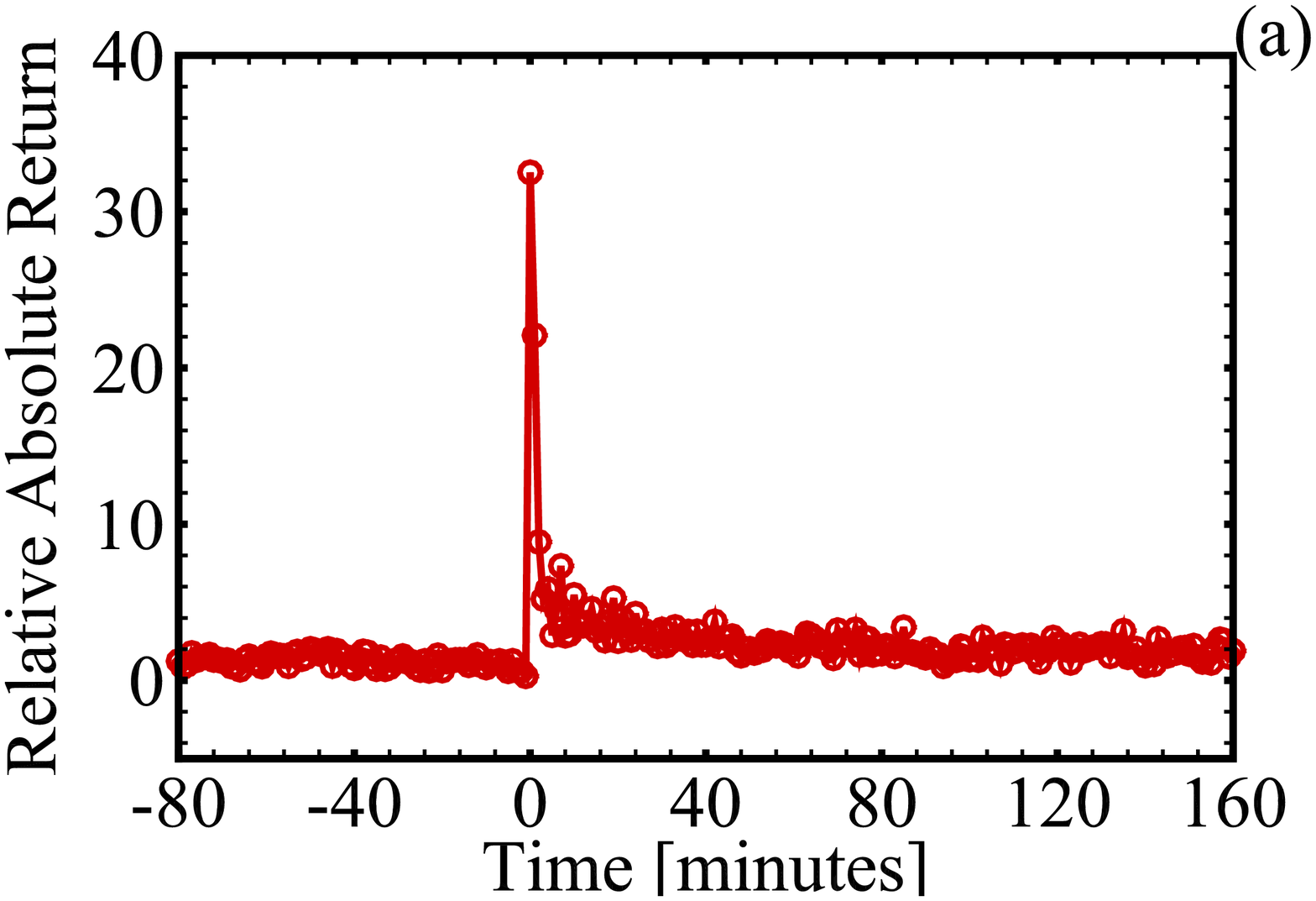}
  \includegraphics[width=0.32\textwidth]{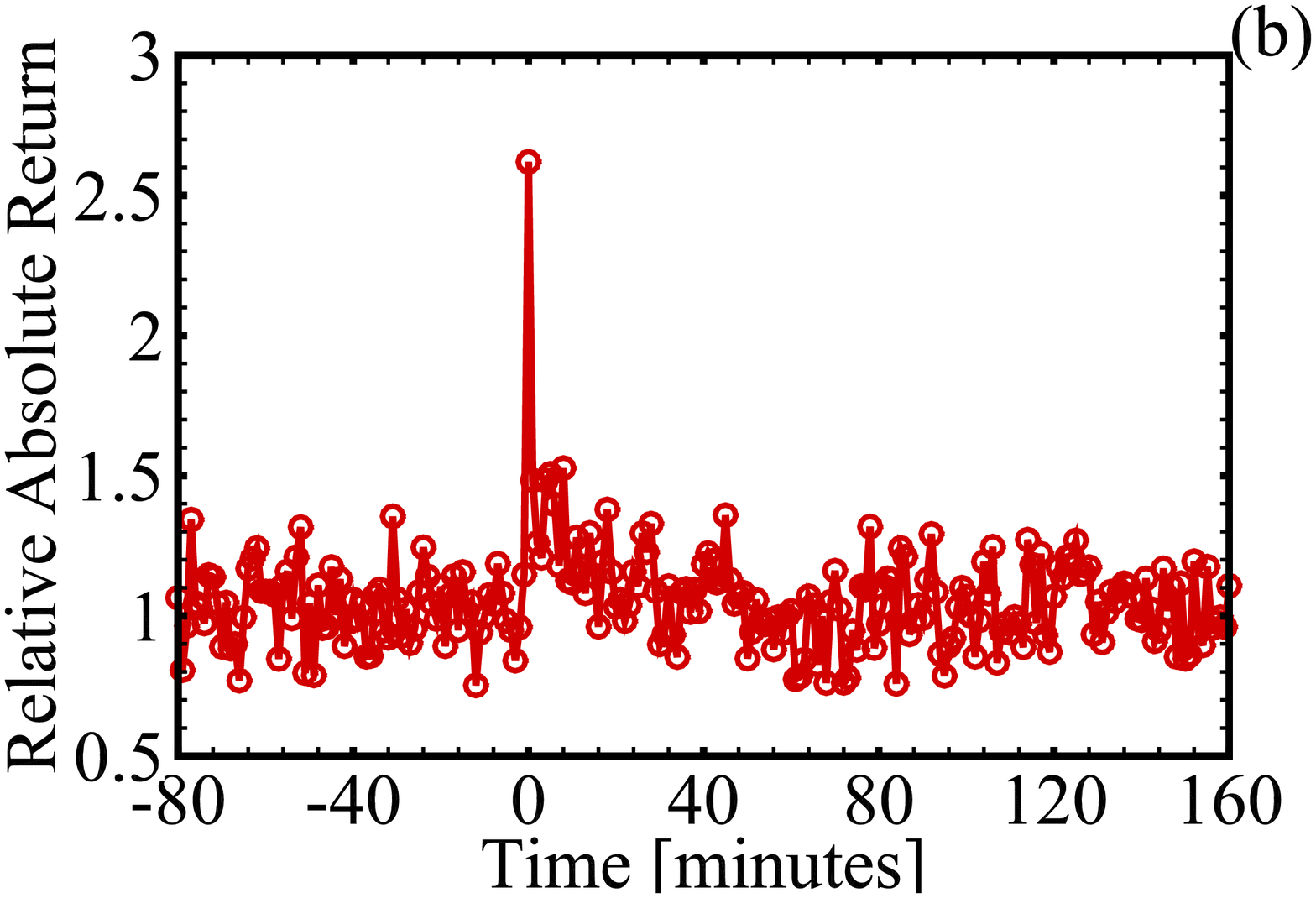}
  \includegraphics[width=0.32\textwidth]{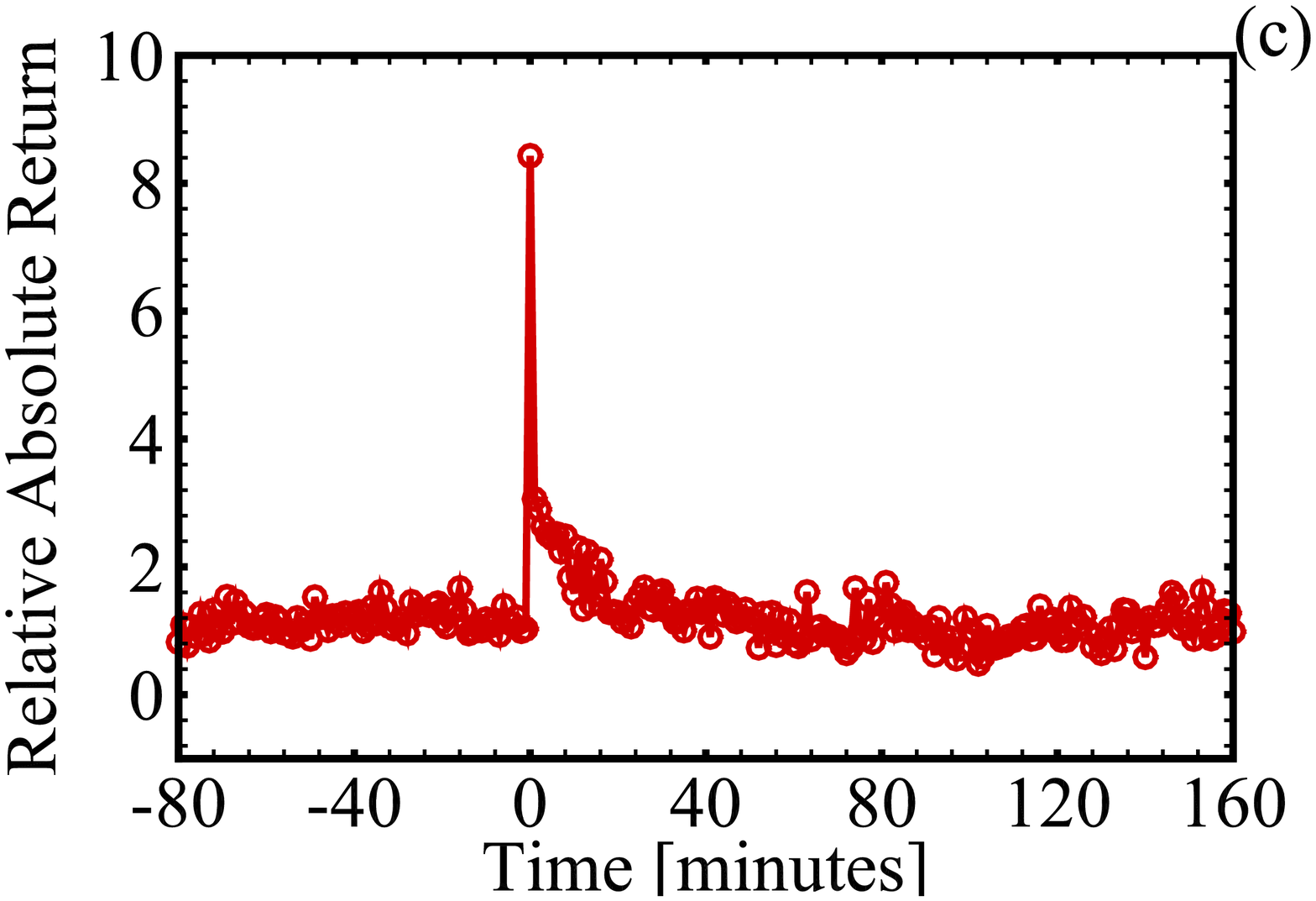}
  \includegraphics[width=0.32\textwidth]{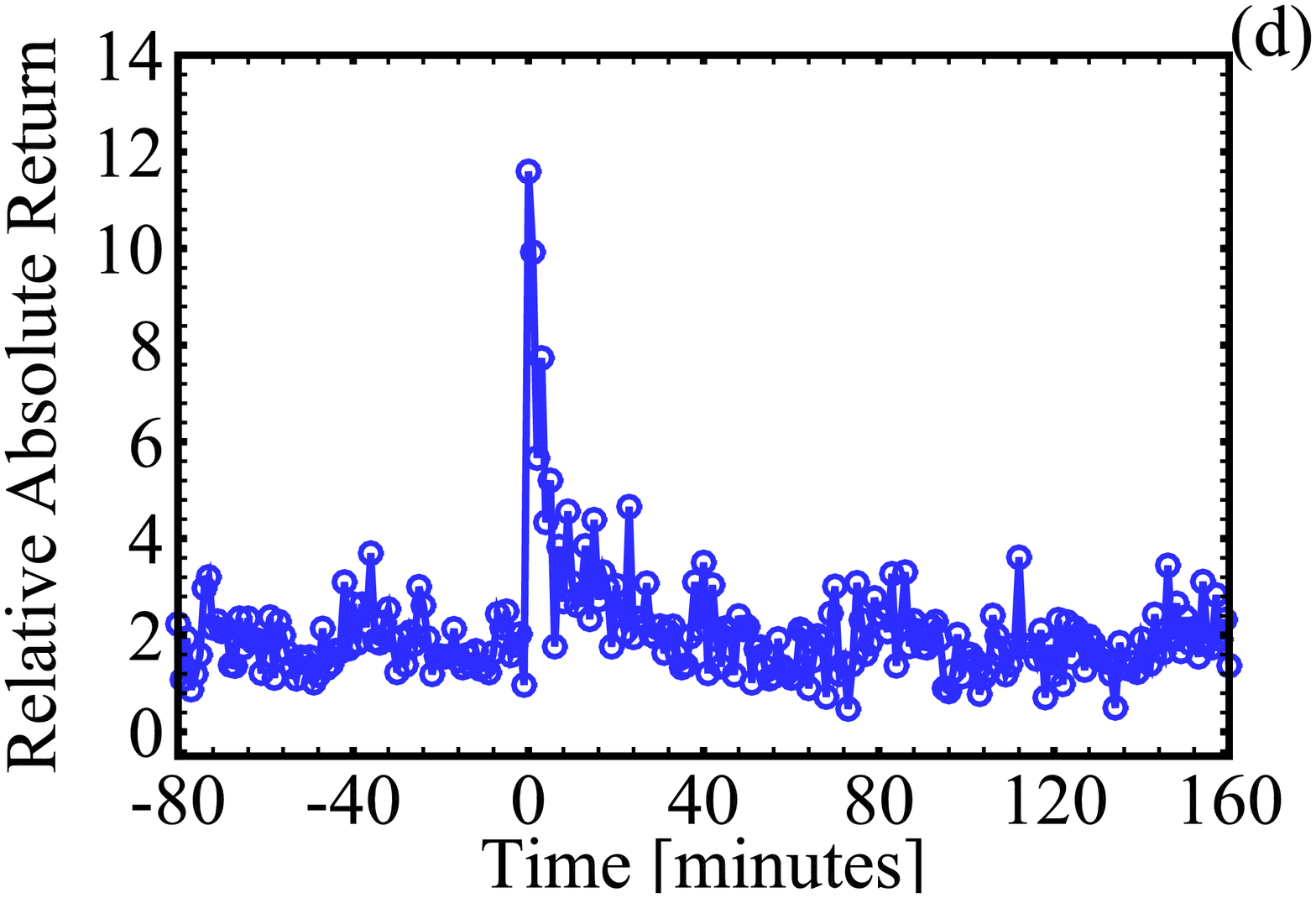}
  \includegraphics[width=0.32\textwidth]{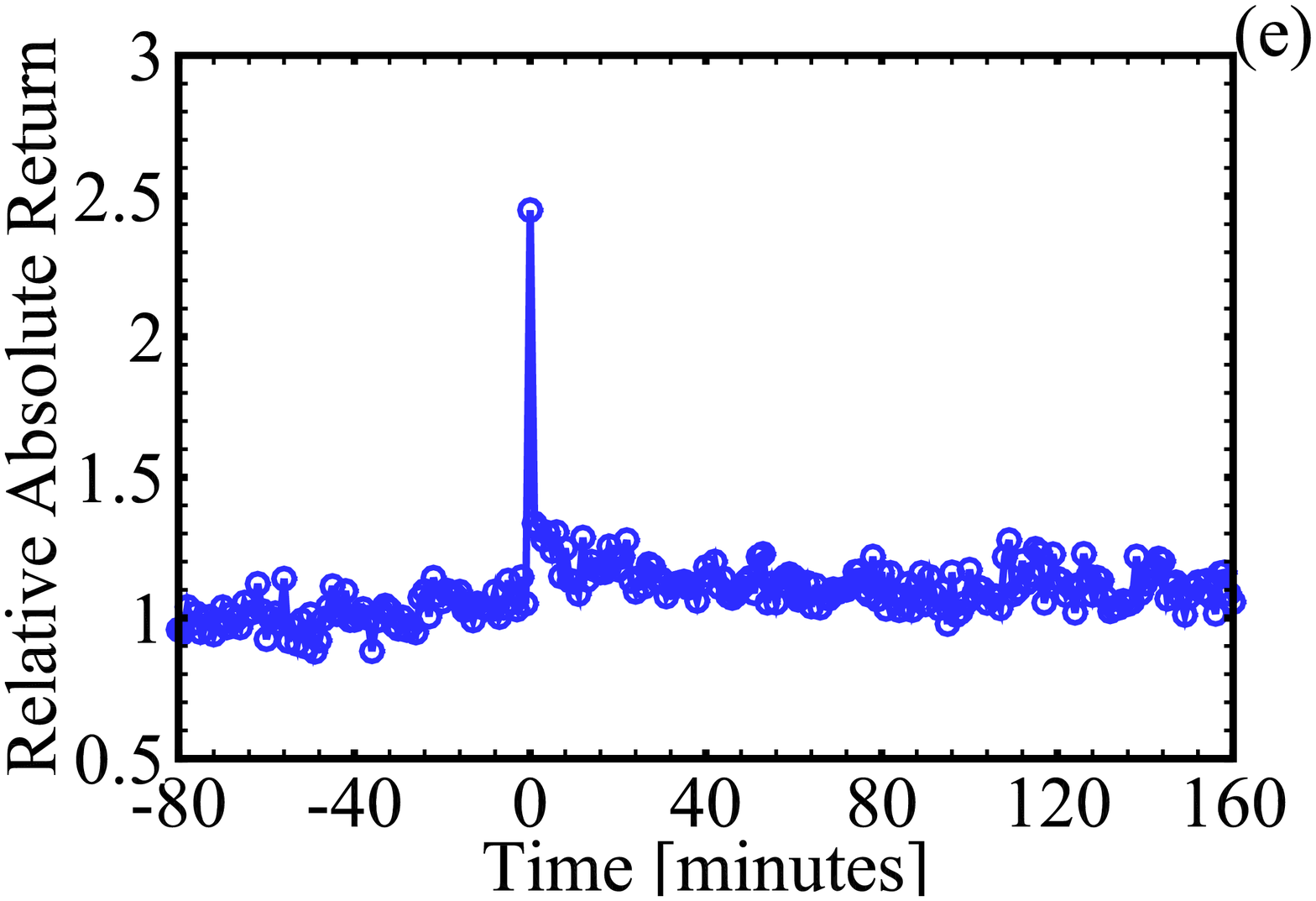}
  \includegraphics[width=0.32\textwidth]{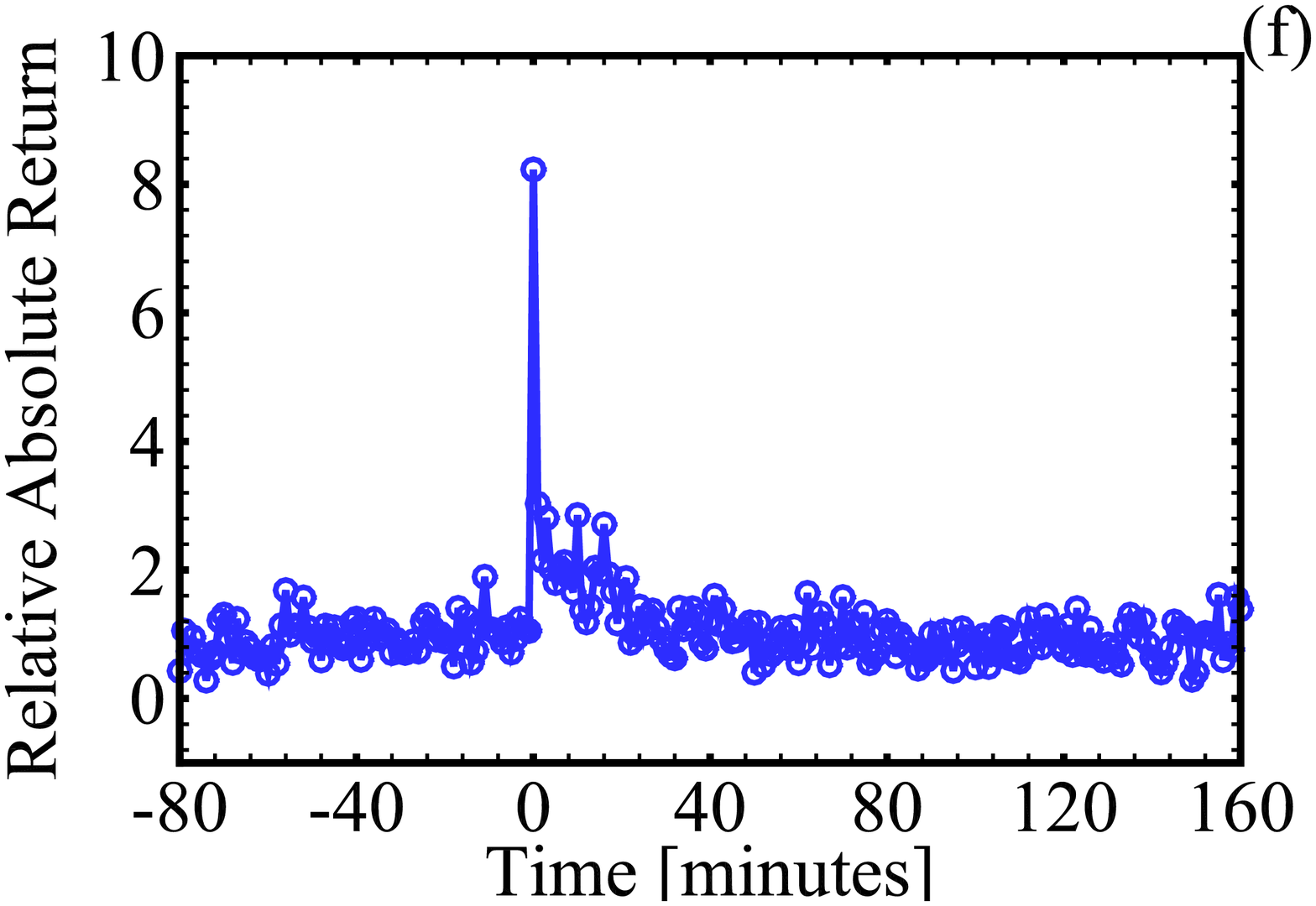}
  \caption{\label{Fig:AbsRet:Each} Dynamics of absolute return around three types of trading halts which are also divided into positive and negative events: (a) intraday halts around 14 positive events, (b) one-day halts around 84 positive events, (c) inter-day halts around 28 positive events, (d) intraday halts around 10 negative events, (e) one-day halts around 489 negative events, and (f) inter-day halts around 15 negative events.}
\end{figure}

\begin{figure}[!htb]
  \centering
  \includegraphics[width=0.32\textwidth]{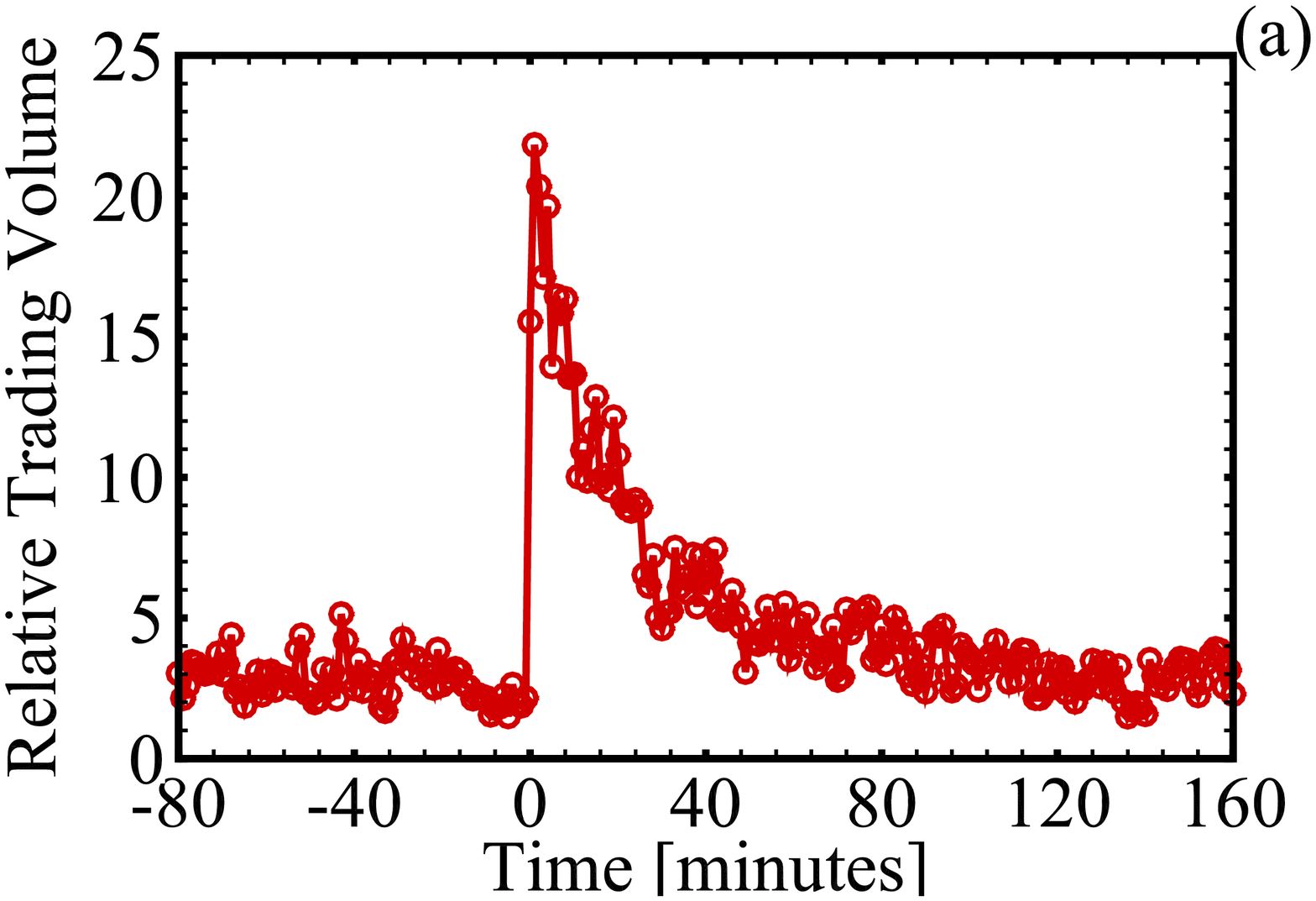}
  \includegraphics[width=0.32\textwidth]{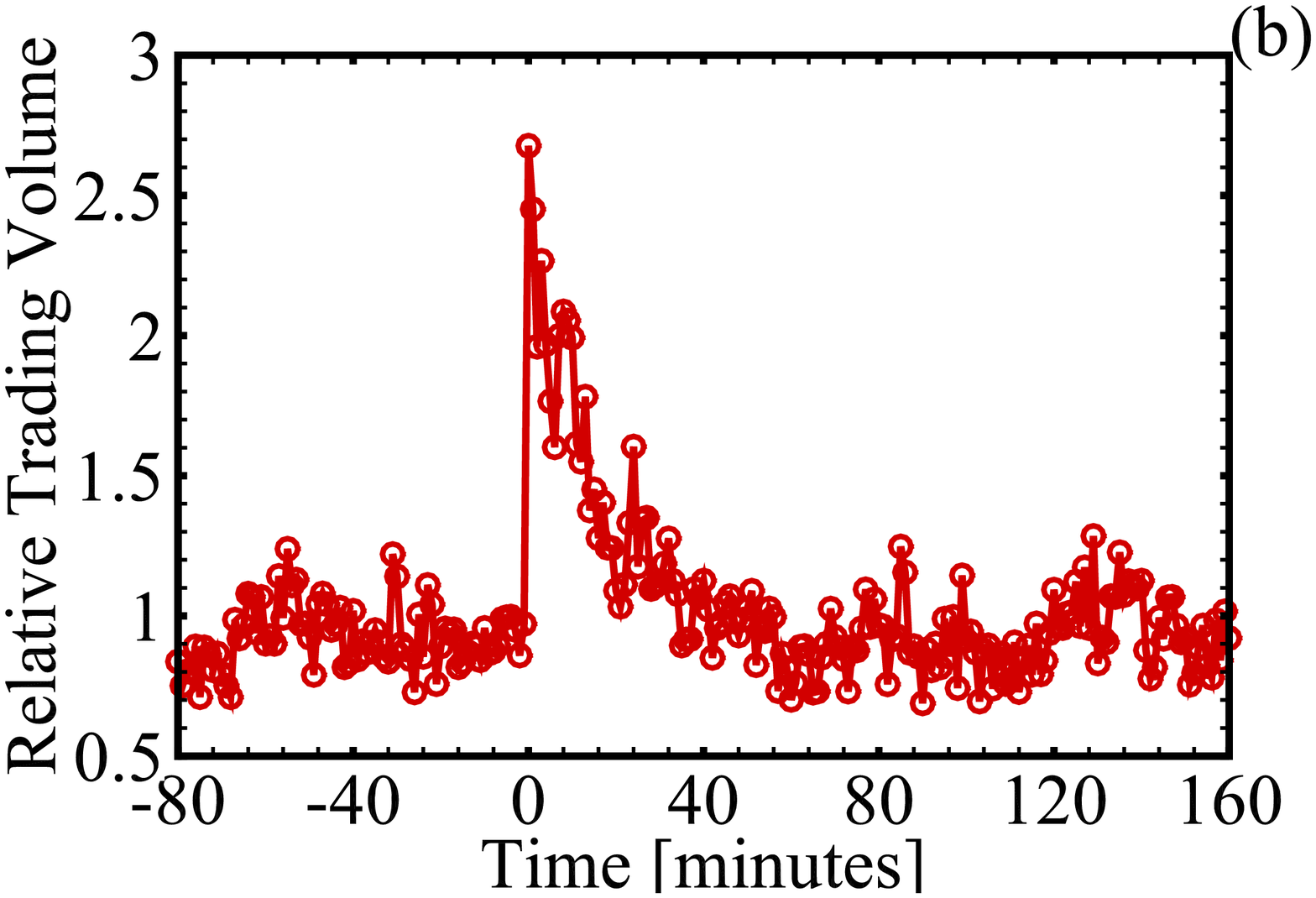}
  \includegraphics[width=0.32\textwidth]{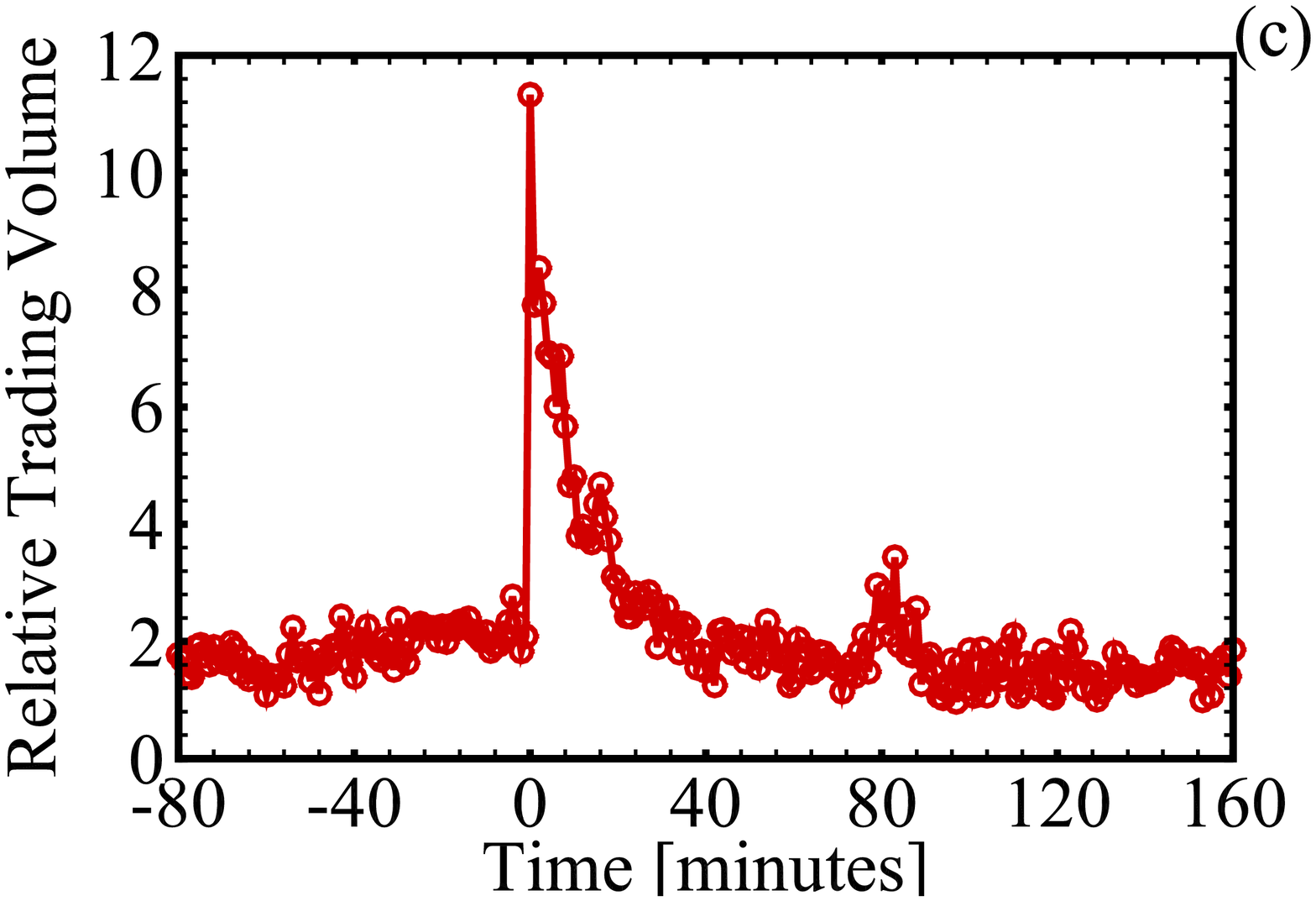}
  \includegraphics[width=0.32\textwidth]{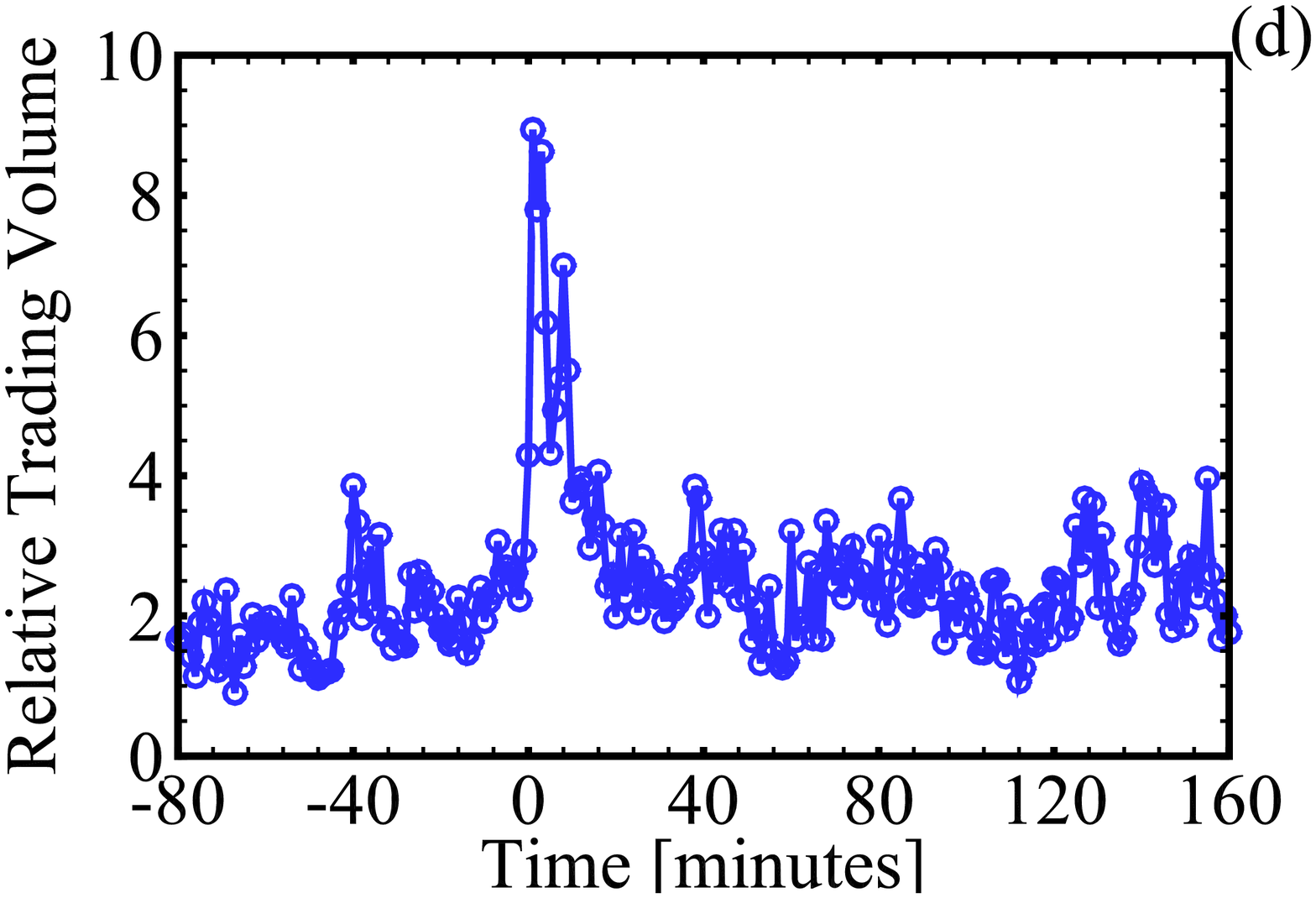}
  \includegraphics[width=0.32\textwidth]{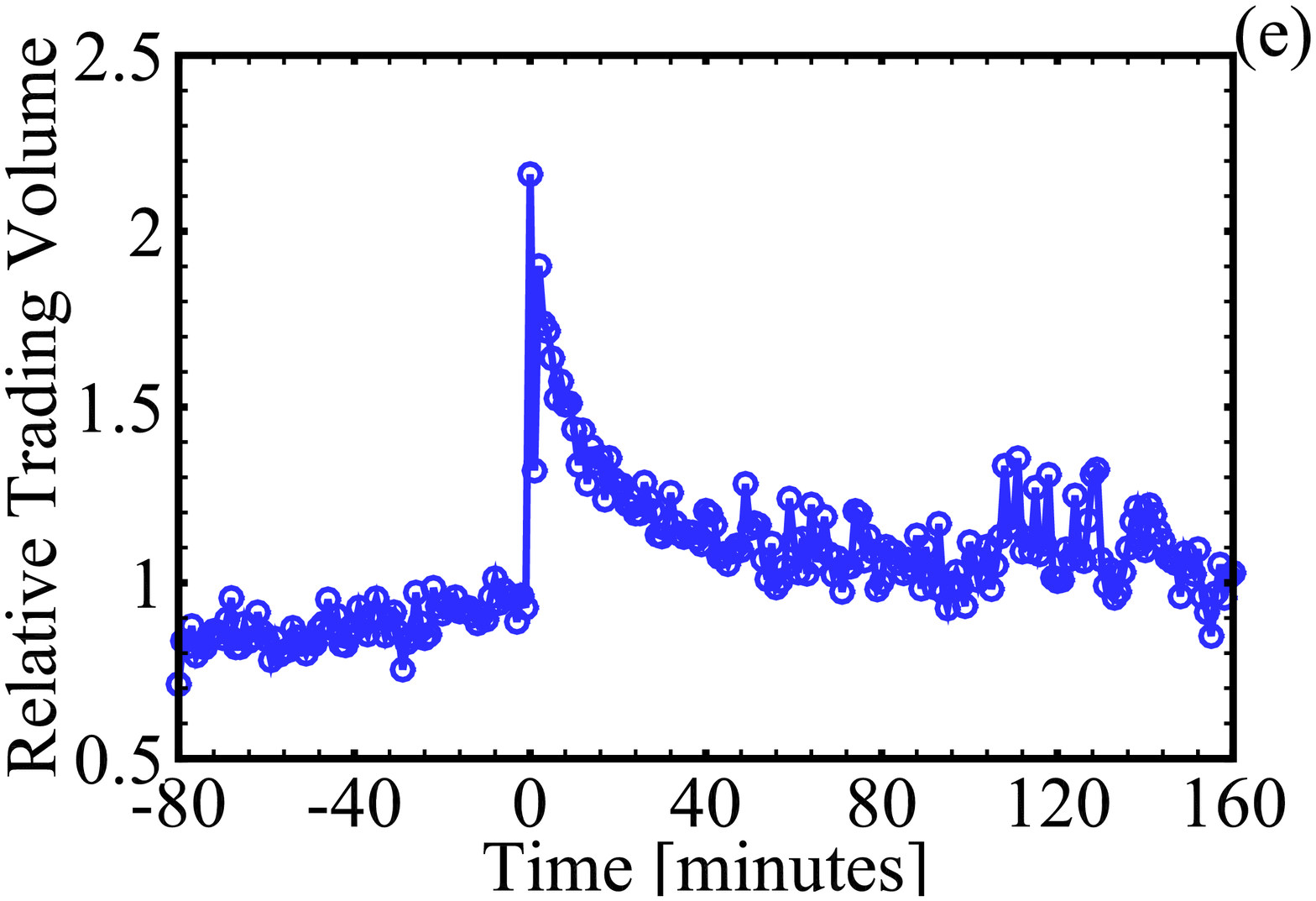}
  \includegraphics[width=0.32\textwidth]{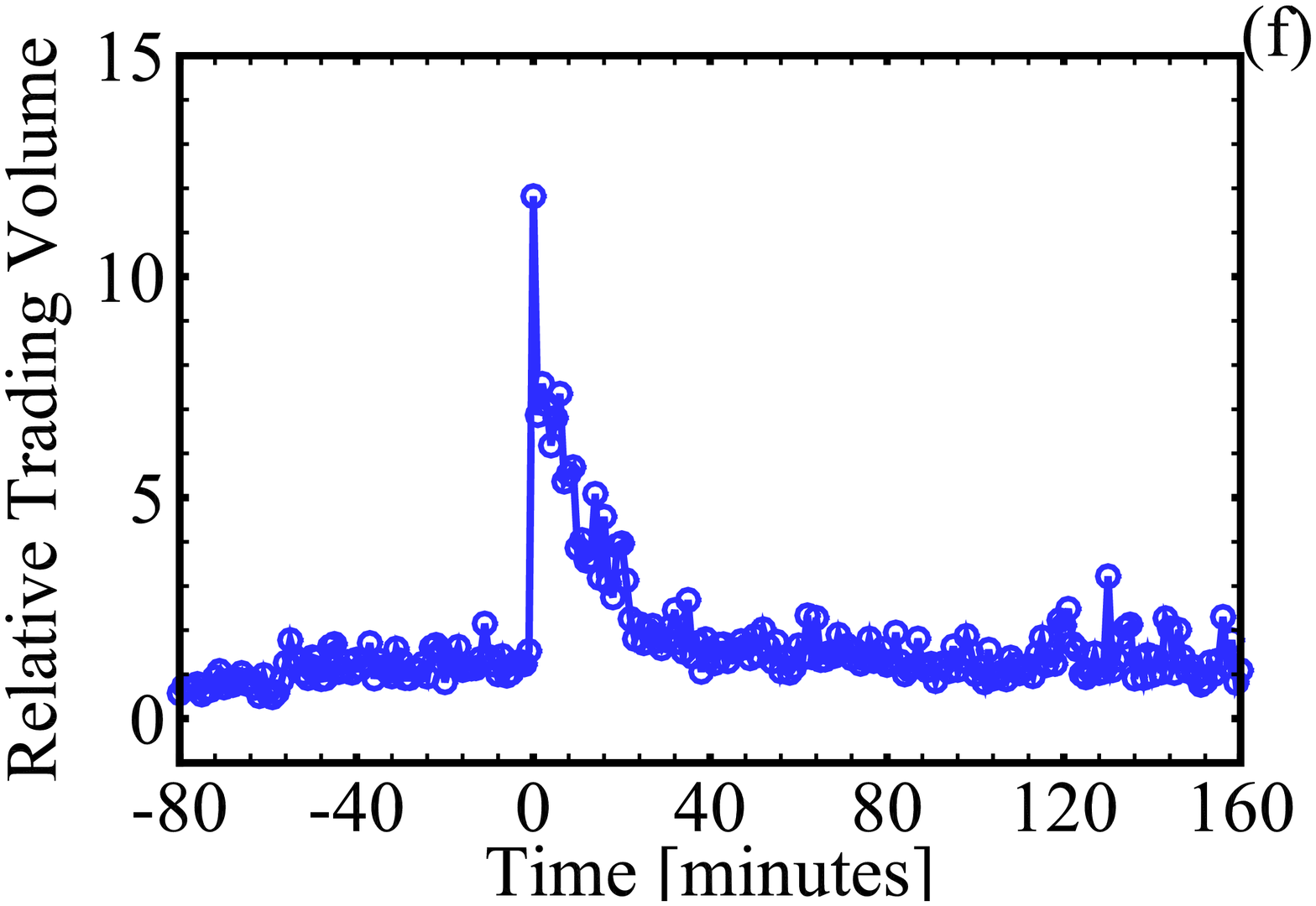}
  \caption{\label{Fig:TraVol:Each} Dynamics of trading volume around three types of trading halts which are also divided into positive and negative events: (a) intraday halts around 14 positive events, (b) one-day halts around 84 positive events, (c) inter-day halts around 28 positive events, (d) intraday halts around 10 negative events, (e) one-day halts around 489 negative events, and (f) inter-day halts around 15 negative events.}
\end{figure}

\begin{figure}[!htb]
  \centering
  \includegraphics[width=0.32\textwidth]{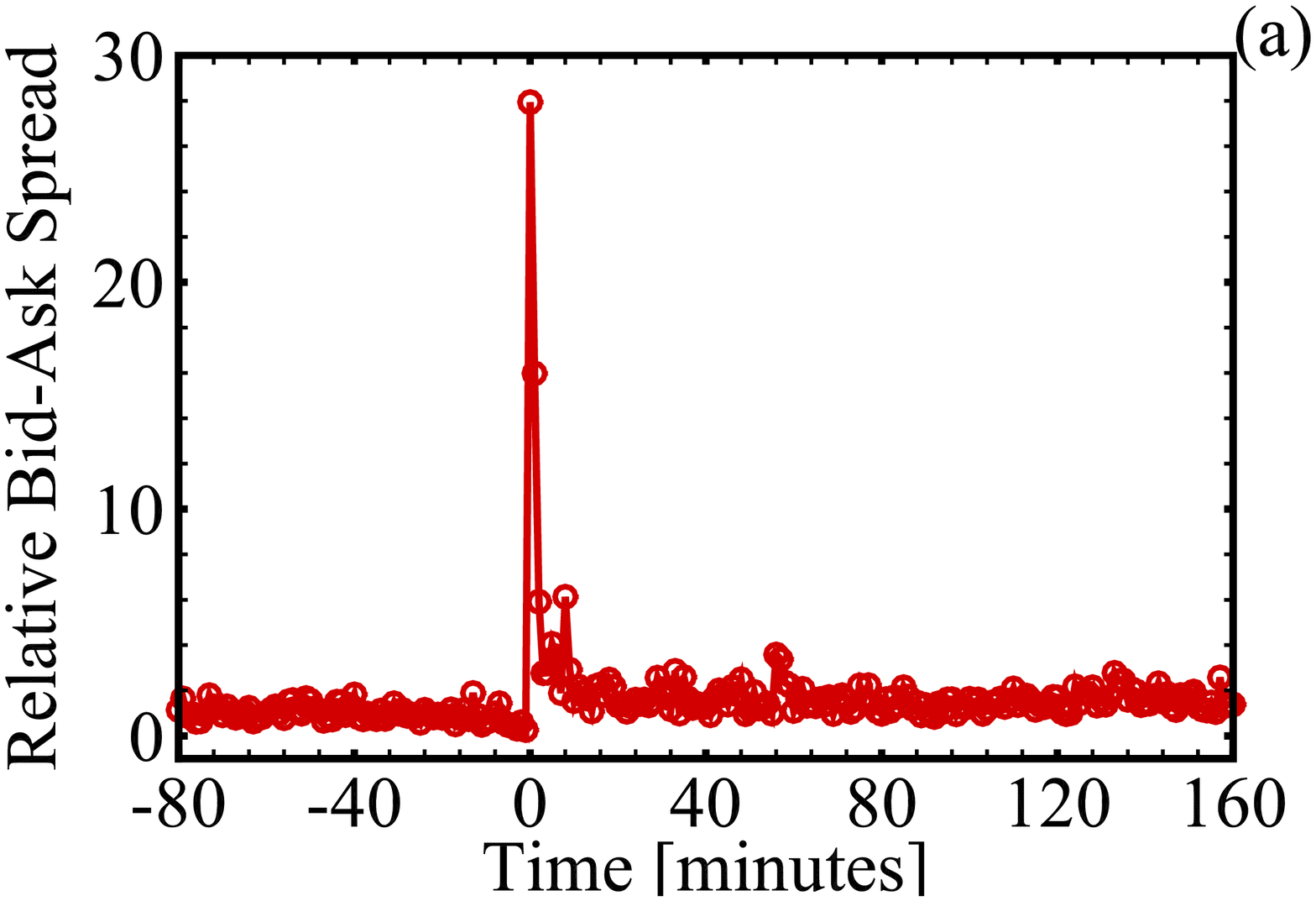}
  \includegraphics[width=0.32\textwidth]{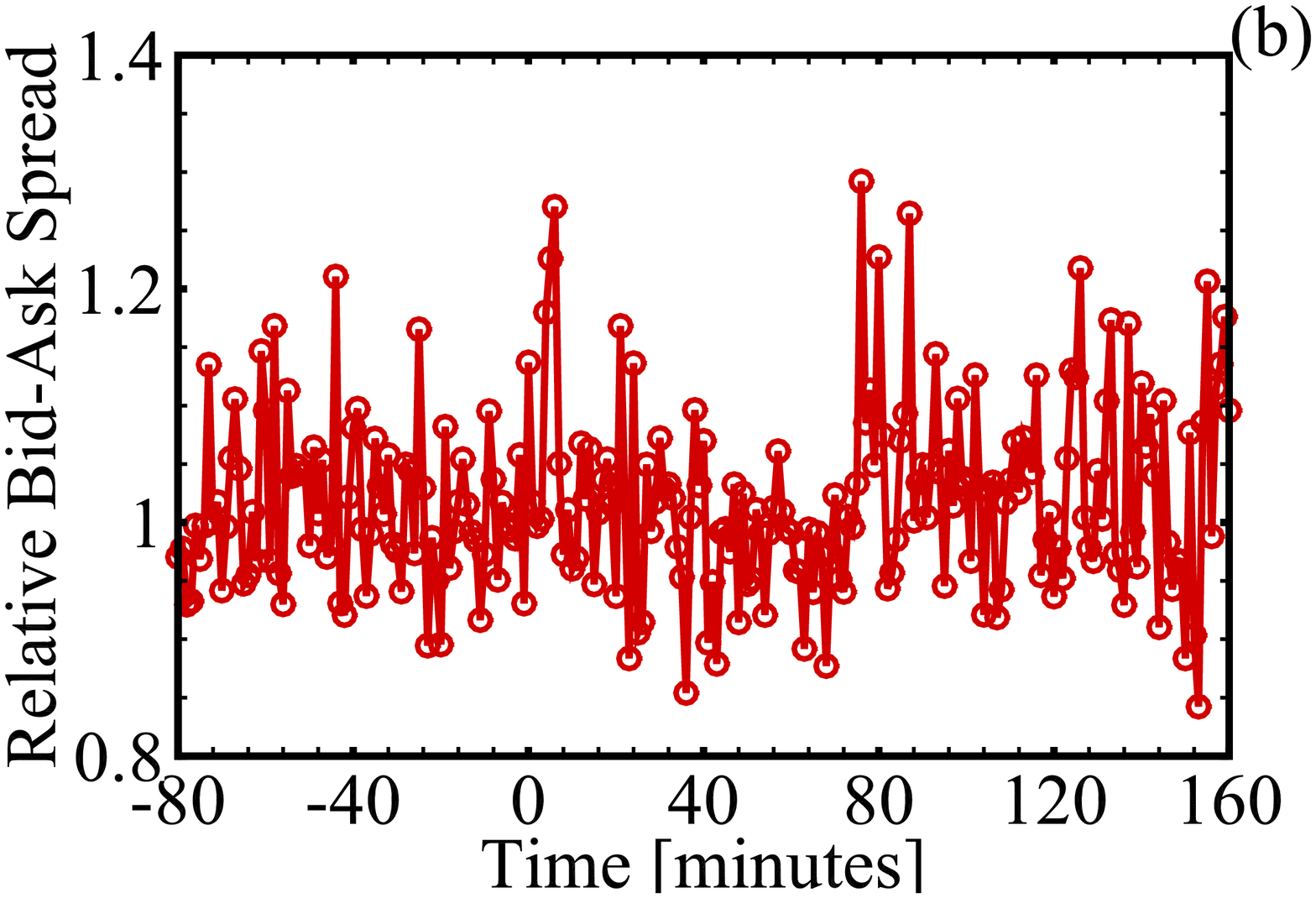}
  \includegraphics[width=0.32\textwidth]{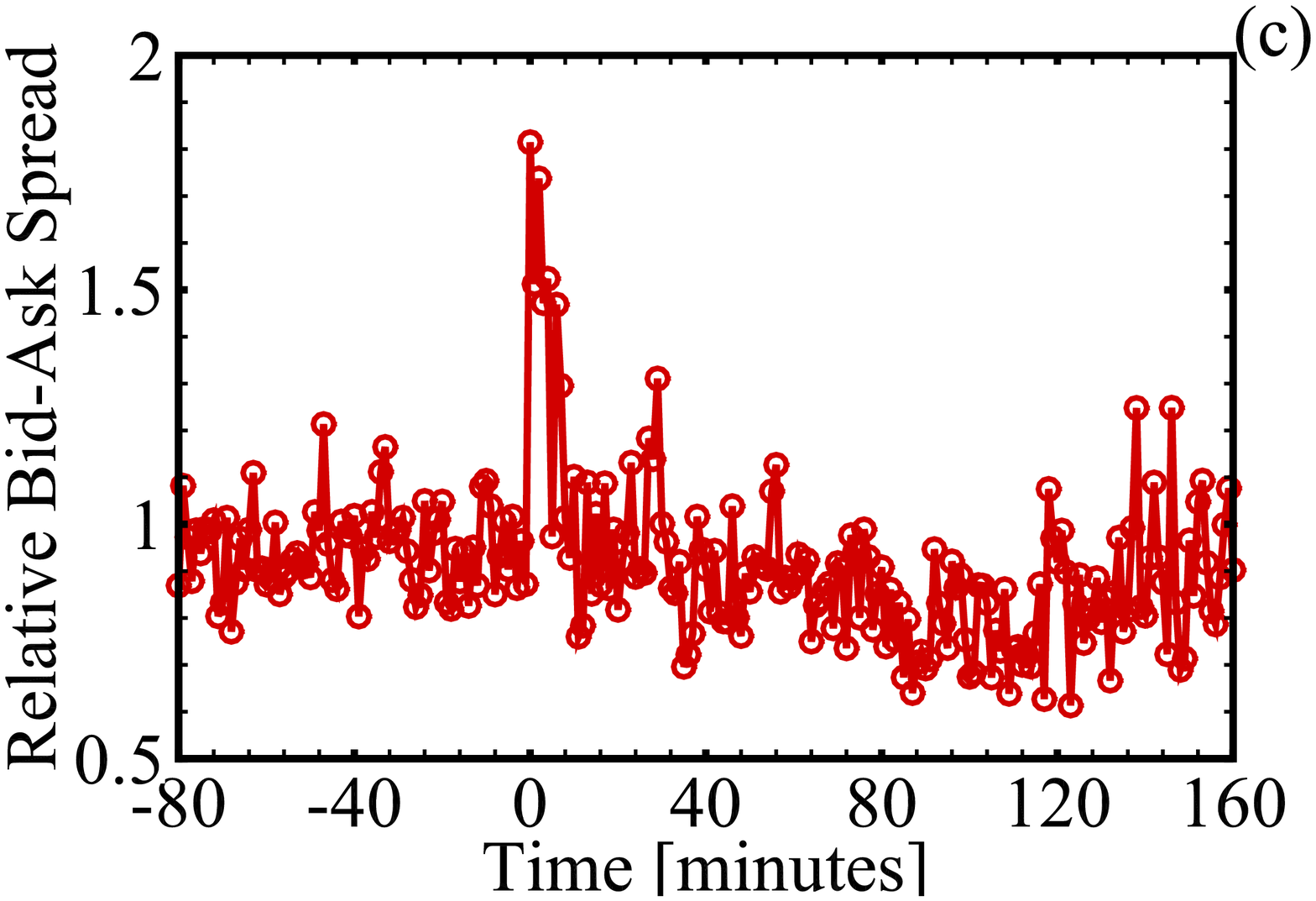}
  \includegraphics[width=0.32\textwidth]{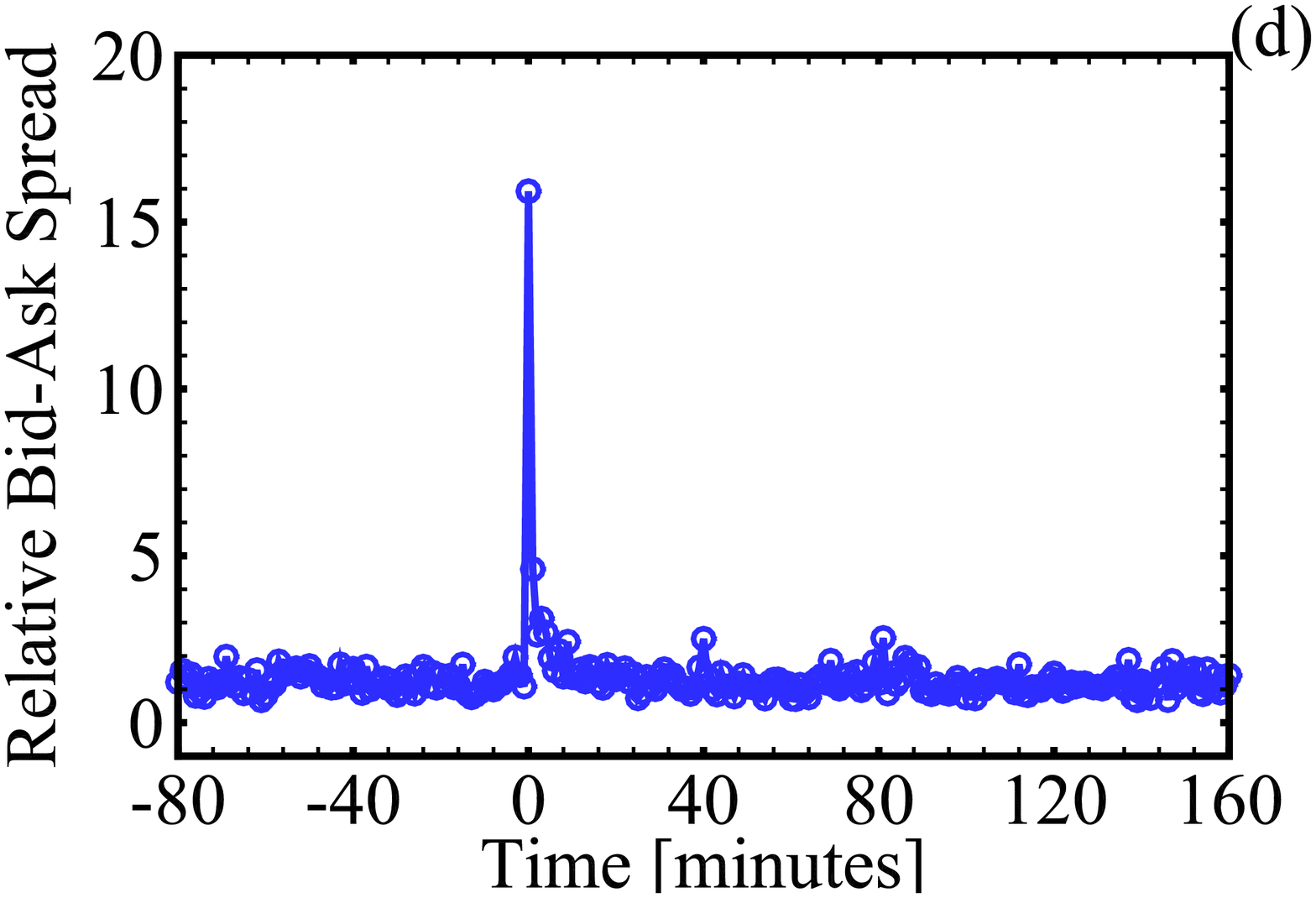}
  \includegraphics[width=0.32\textwidth]{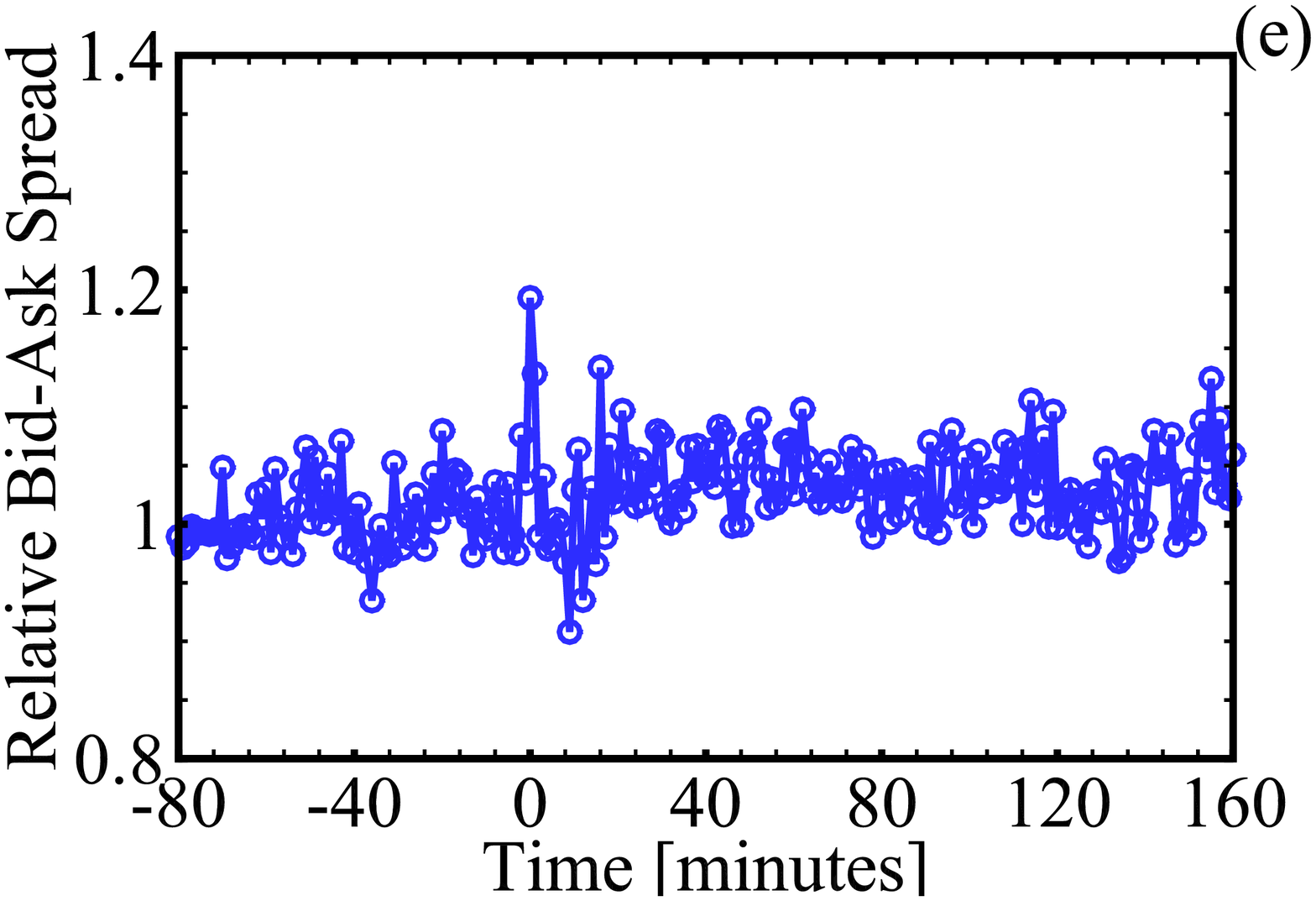}
  \includegraphics[width=0.32\textwidth]{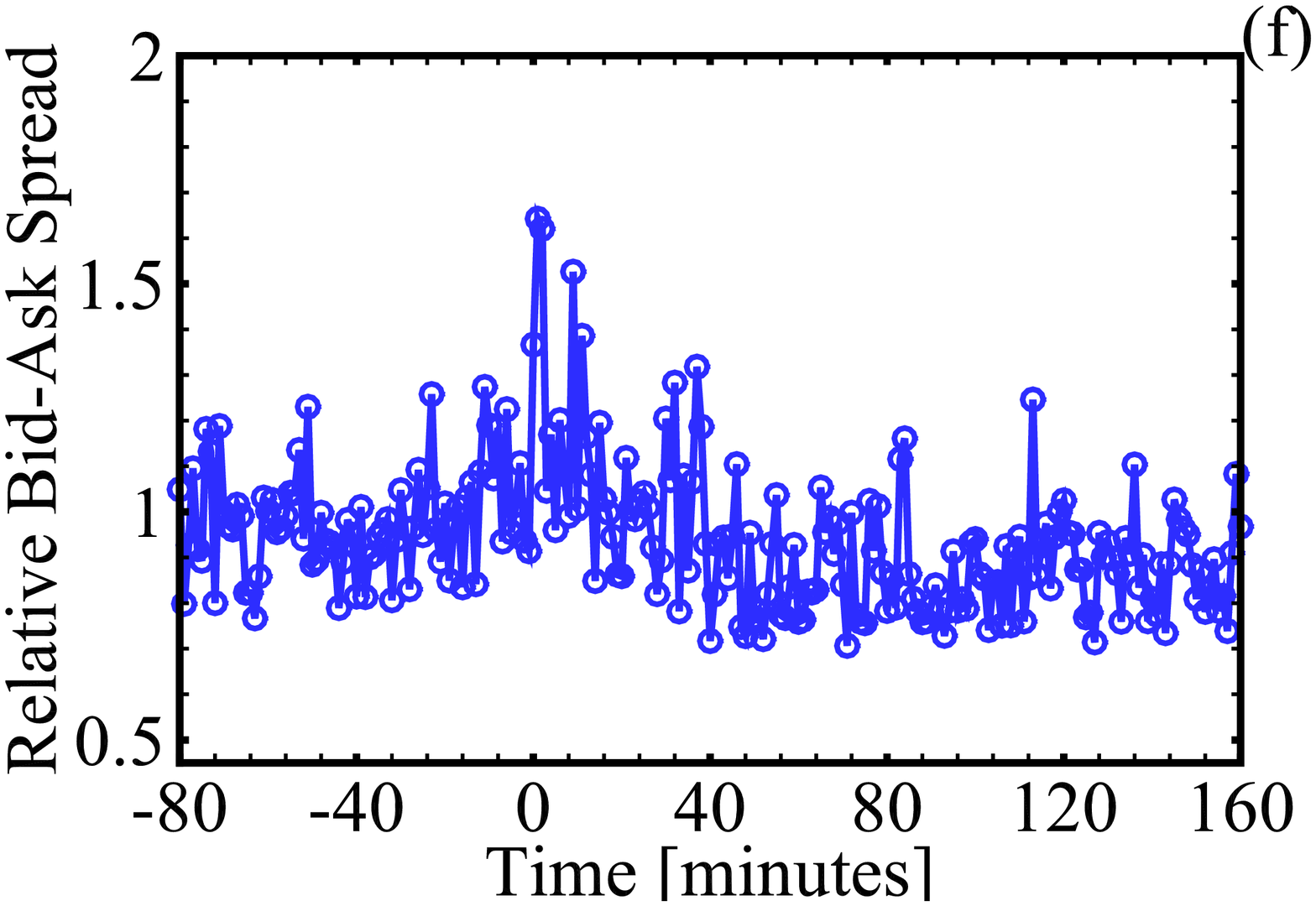}
  \caption{\label{Fig:BidAskSpr:Each} Dynamics of bid-ask spread around three types of trading halts which are also divided into positive and negative events: (a) intraday halts around 14 positive events, (b) one-day halts around 84 positive events, (c) inter-day halts around 28 positive events, (d) intraday halts around 10 negative events, (e) one-day halts around 489 negative events, and (f) inter-day halts around 15 negative events.}
\end{figure}

Figures~\ref{Fig:AbsRet:Each}(a)-~\ref{Fig:AbsRet:Each}(f) give the average evolution of absolute return around 3 types of trading halts, which also are divided into positive and negative events. It can be found that all six plots share the same pattern with a sharp peak at $t_{max}=\rm{0}$ and a suddenly increasing before trading halts and a relatively rapid decay after that, which is very similar with the volatility analysis of microtrends switches and macrotrends switches in Refs. \cite{Preis-Stanley-2010-JSP}. Actually, trading halts are natural exogenous shocks, the typical signatures of which are the sudden peak and relatively rapid relaxation \cite{Crane-Sornette-2008-PNAS}. In addition, one obvious feature for both positive events and negative events is that the peak of intraday halts is much higher than that of one-day halts and inter-day halts. This is related to the properties of different types of trading halts. The intraday halts are due to abnormal price fluctuations and trading halts further magnify the fluctuation rather than lower the fluctuation. This may look strange but do consist with the trading halt model of Subrahmanyam \cite{Subrahmanyam-1994-JF}. While one-day halts are due to shareholders\textquoteright{} meeting, which is routine halts, so it is rational that one-day halts have lowest peak even considering the magnified fluctuation. The peak of inter-day halts due to announcement of significant events is between that of intraday halts and one-day halts. Another feature for these six plots is that the peaks of positive events are all higher than those of negative events. This indicates that halts can cause greater instant fluctuation for positive events compared with negative events. Furthermore, it should be stressed that intraday halts show big different peaks between positive ones and negative ones, while it is not so obvious for other types of halts. This can be explained by halts' property and traders' irrational behavior. Intraday halts disclose information no more than the price fluctuations which are very uncertain signals for traders. Meanwhile, traders behave differently for positive signals and negative signals in an uncertain environment. More specifically, traders are more willing to accept there is good news behind abnormal price increase, while they are reluctant to accept there is bad news behind abnormal price decline. This asymmetric psychology leads traders to be overconfident when halt is positive, thus to be urgent to deal and increase trading activity, finally lead to larger price changes (higher peak); in contrary, when halt is negative, traders with loss aversion and hesitation lead to lower price changes (lower peak) compared with positive halt's. On the other hand, for one-day halts and intra-day halts, it is hardly to show big different peaks between positive ones and negative ones because these two types halts disclose more certain information and then traders behave less asymmetrically.

Figures~\ref{Fig:TraVol:Each}(a)-~\ref{Fig:TraVol:Each}(f) show the average evolution of trading volume around 3 types of trading halts. We can find these plots have similar peaks with above absolute return plots, but have more discernible decay and slower relaxation after trading halts. The asymmetric volume dynamics before and after trading halts illustrate the response feature of exogenous events \cite{Crane-Sornette-2008-PNAS} again. In case of positive events, consistent with the evolution charts of absolute return, intraday halts show the highest peak, and one-day halts show the lowest peak in the volume dynamics. While in case of negative events, intraday halts do not display an abnormally high peak like the positive intraday halts, which may be explained as the result of confused future about the stock price under the condition of large price fall, thus investors trade prudentially even after trading halt.

On the other hand, in case of the bid-ask spread (Figure~\ref{Fig:BidAskSpr:Each}), which is a major source of transaction costs, we find that intraday halts show the sharp peak pattern and a fast decay which is consistent with the empirical result of trading halts on Nasdaq \cite{Christie-Corwin-Harris-2002-JF}. However, inter-day halts display relatively lower peak and practically no significant change of bid-ask spread accompanies the one-day halts. The reason why bid-ask spread enlargement does not occur in case of inter-day halts can be attributed to the information content of different halts. Information disclosed by intraday halts is much vaguer than by inter-day halts, which announce significant events. Consequently, larger bid-ask spread reflects higher information cost after intraday halts. In contrast, due to more explicit information provided by inter-day halts, the bid-ask spread enlargements are not obvious. Meanwhile, this implies that a contrarian strategy following the intraday halt has higher costs than following one-day halt or inter-day halt. In addition, consistent with absolute return and trading volume above, the peak of positive events is higher than that of negative events in bid-ask spread dynamics of intraday halts. Such difference can also be explained by traders' irrationality as before, that is, traders are more willing to accept there is good news behind abnormal price increase, while they are reluctant to accept there is bad news behind abnormal price decline. This asymmetric psychology leads traders to be overconfident when halt is positive, thus traders submit more market orders which consume liquidity and enlarge bid-ask spread significantly; in contrary, when halt is negative, traders with loss aversion and hesitation submit not that many market orders and thus bid-ask spread will not show that high peak.

\section{Power law behavior after trading halts}
\label{S1:PowerLaw}

\begin{figure}[!htb]
  \centering
  \includegraphics[width=0.32\textwidth]{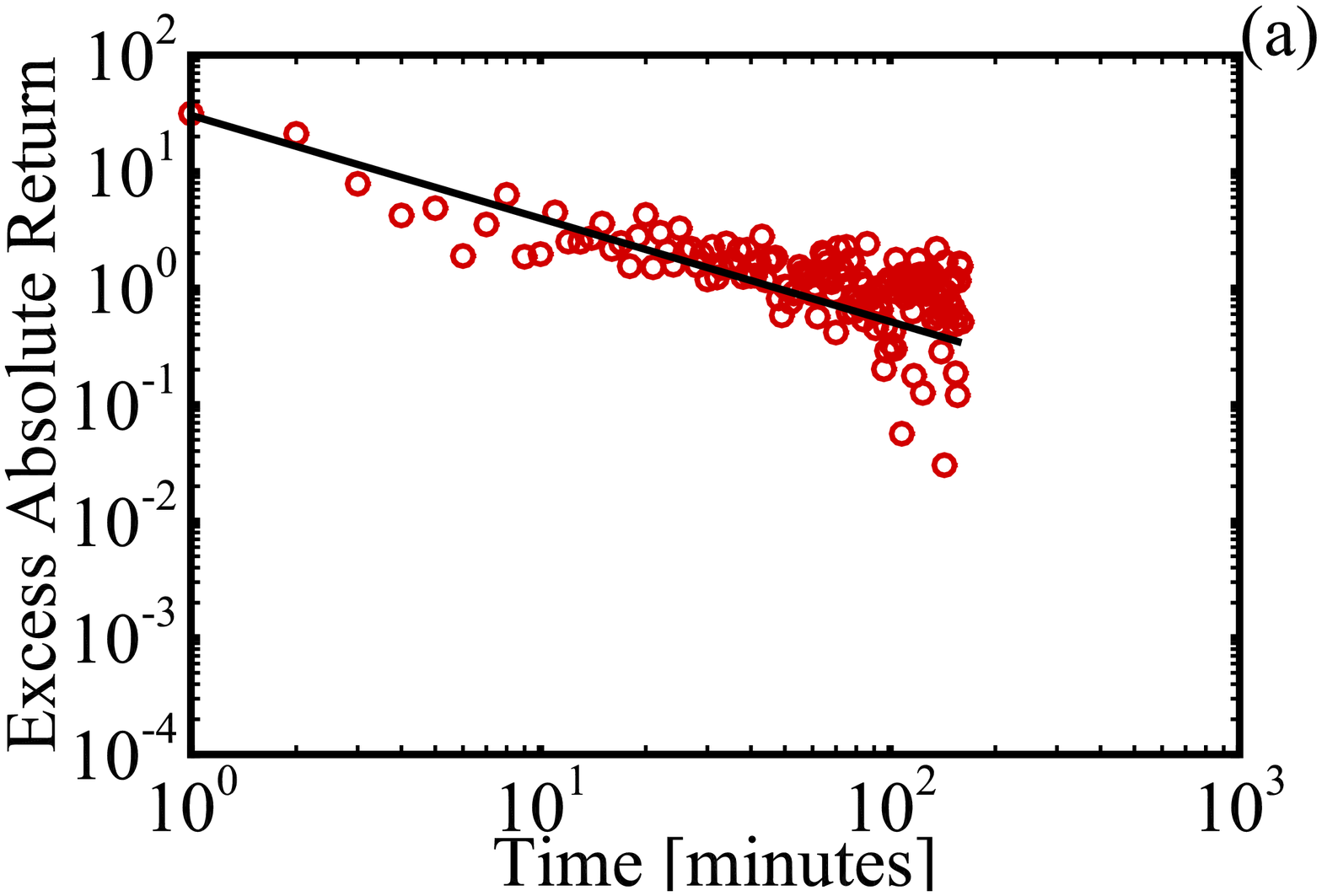}
  \includegraphics[width=0.32\textwidth]{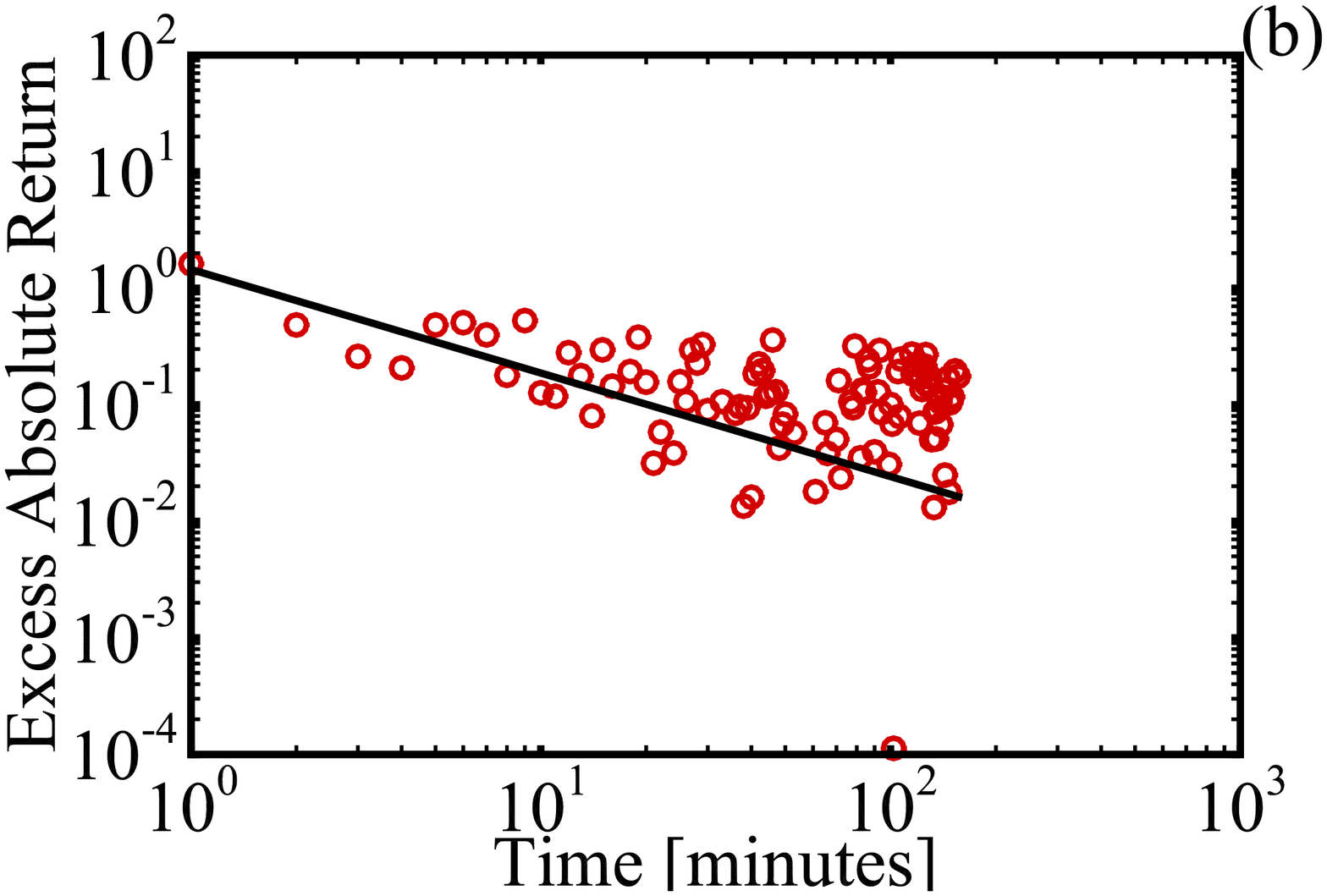}
  \includegraphics[width=0.32\textwidth]{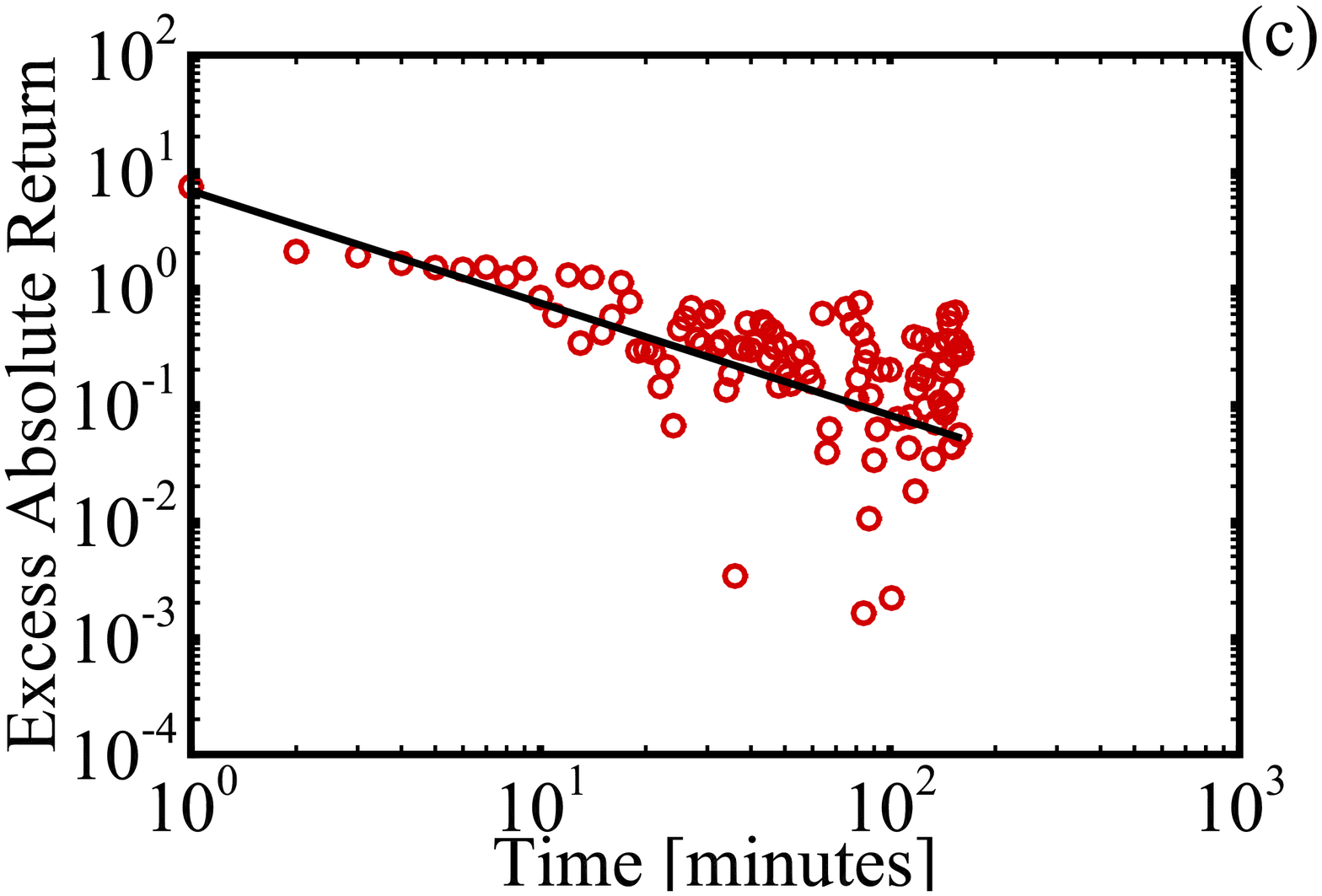}
  \includegraphics[width=0.32\textwidth]{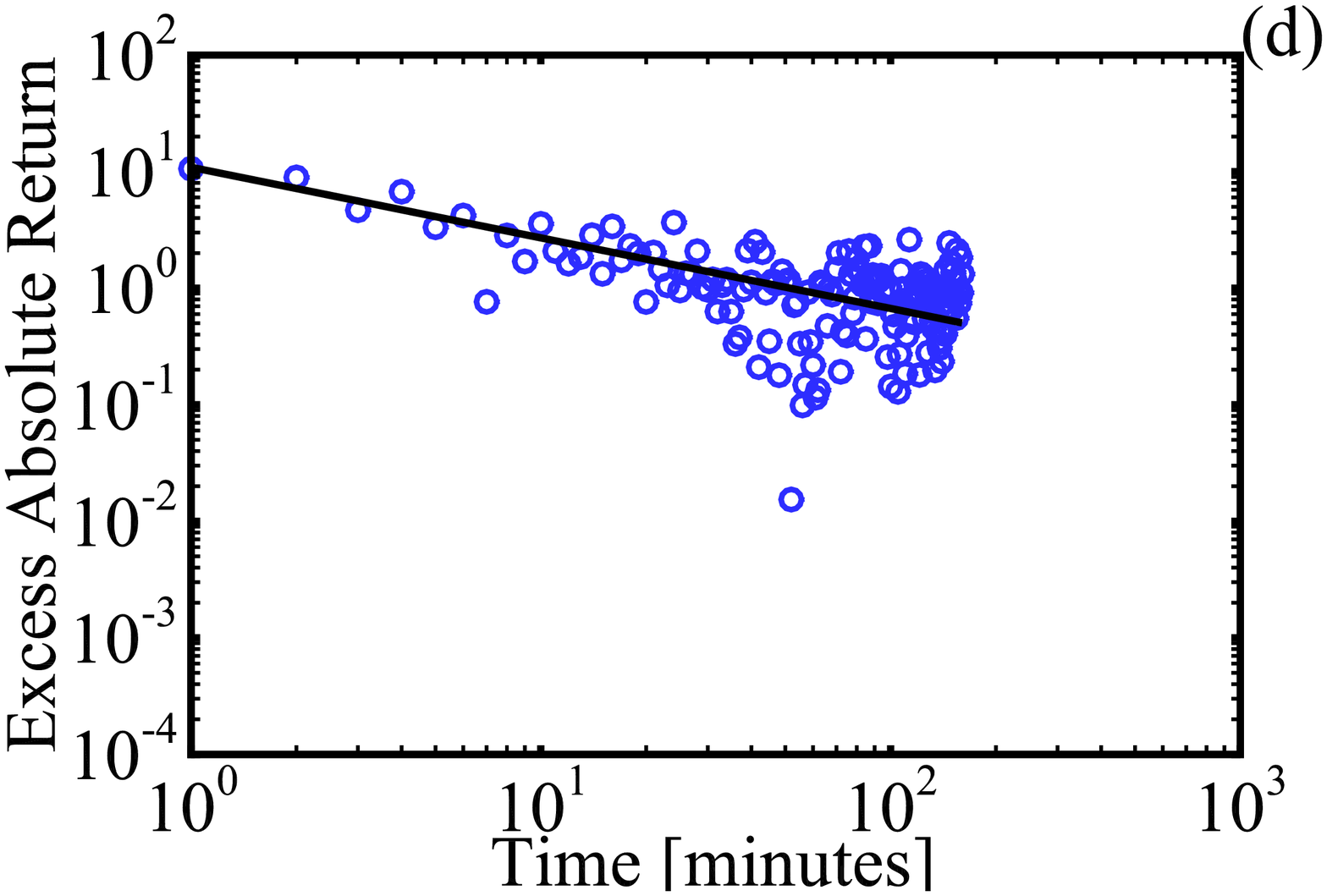}
  \includegraphics[width=0.32\textwidth]{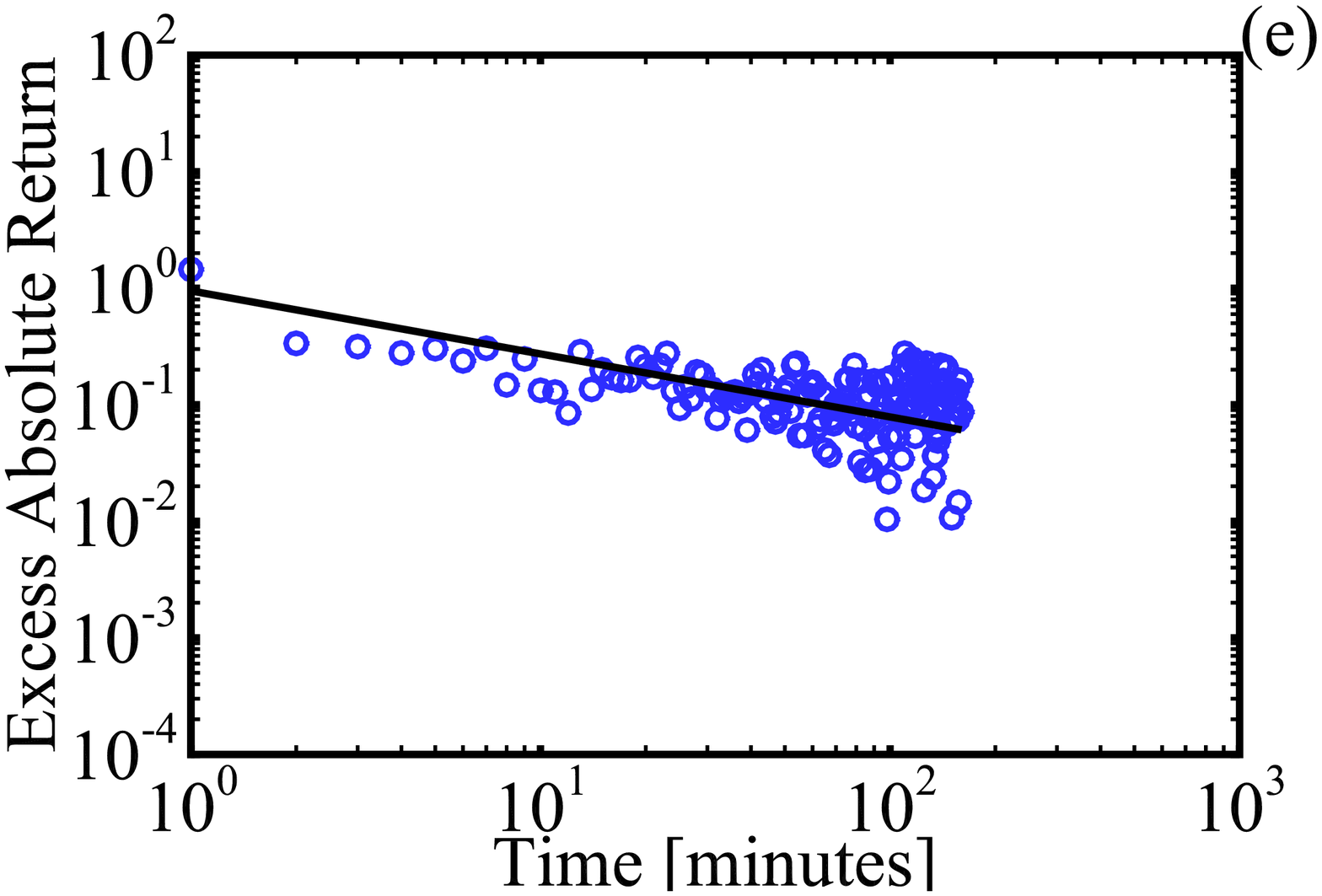}
  \includegraphics[width=0.32\textwidth]{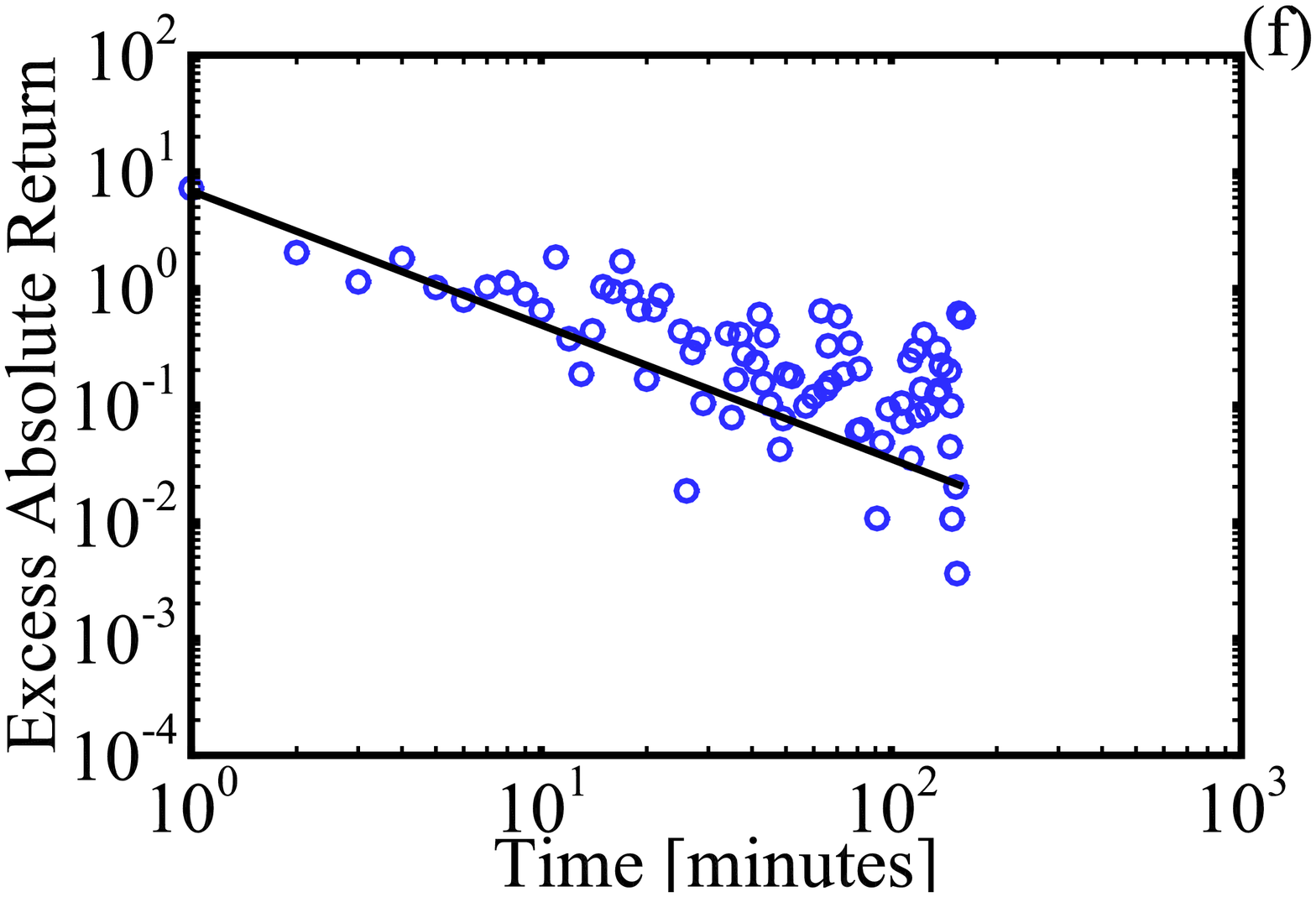}
  \caption{\label{Fig:PowerLaw:ExAbsRet:Each} Power-law relaxation of excess variable for absolute return after three types of trading halts which are also divided into positive and negative events on log-log scales with power-law fits. The relaxation exponents $\alpha$ are: (a) 0.89 $\pm$ 0.05, (b) 0.89 $\pm$ 0.13, (c) 0.96 $\pm$ 0.07, (d) 0.61 $\pm$ 0.05, (e) 0.54 $\pm$ 0.05, and (f) 1.15 $\pm$ 0.13, respectively.}
\end{figure}

\begin{figure}[!htb]
  \centering
  \includegraphics[width=0.32\textwidth]{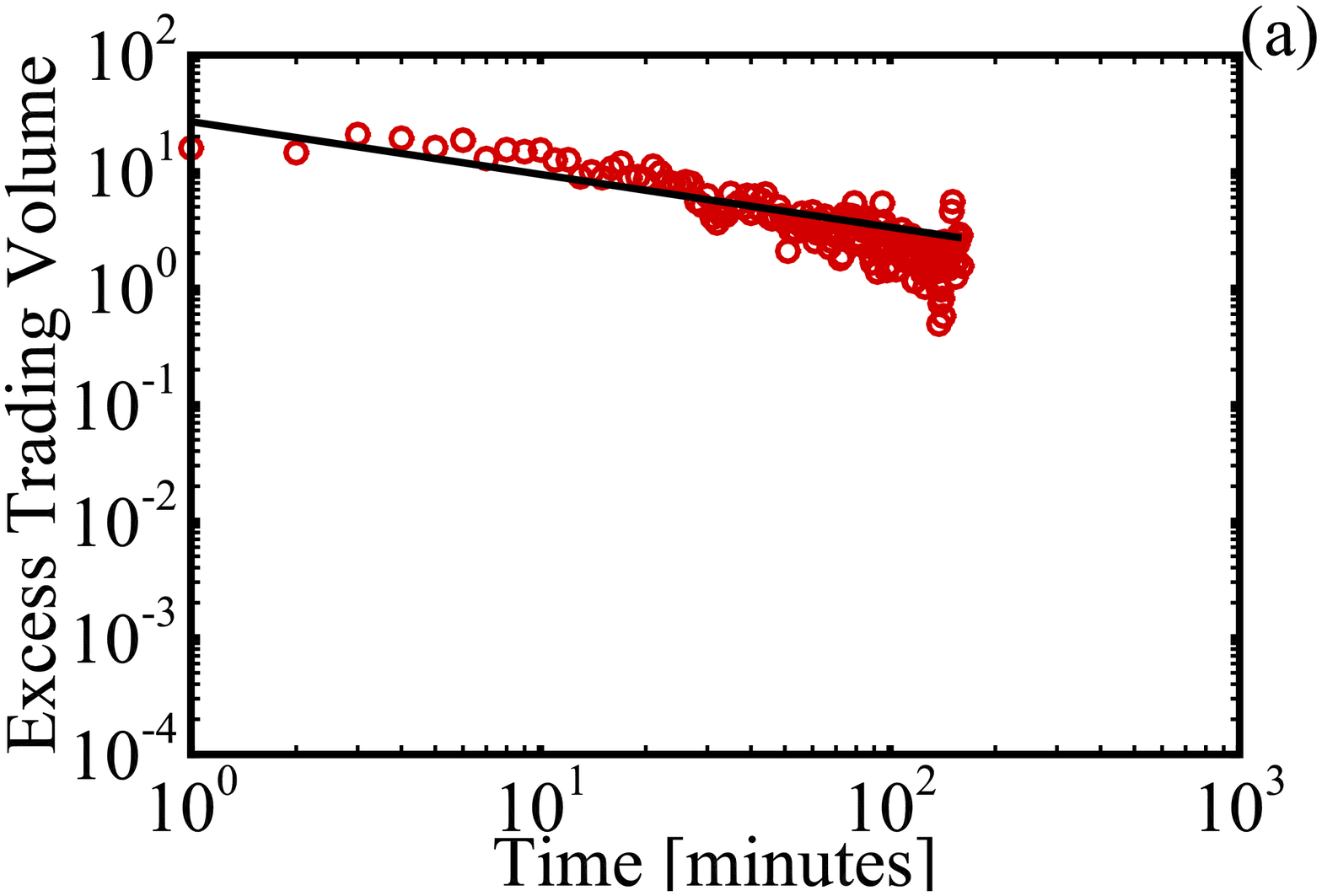}
  \includegraphics[width=0.32\textwidth]{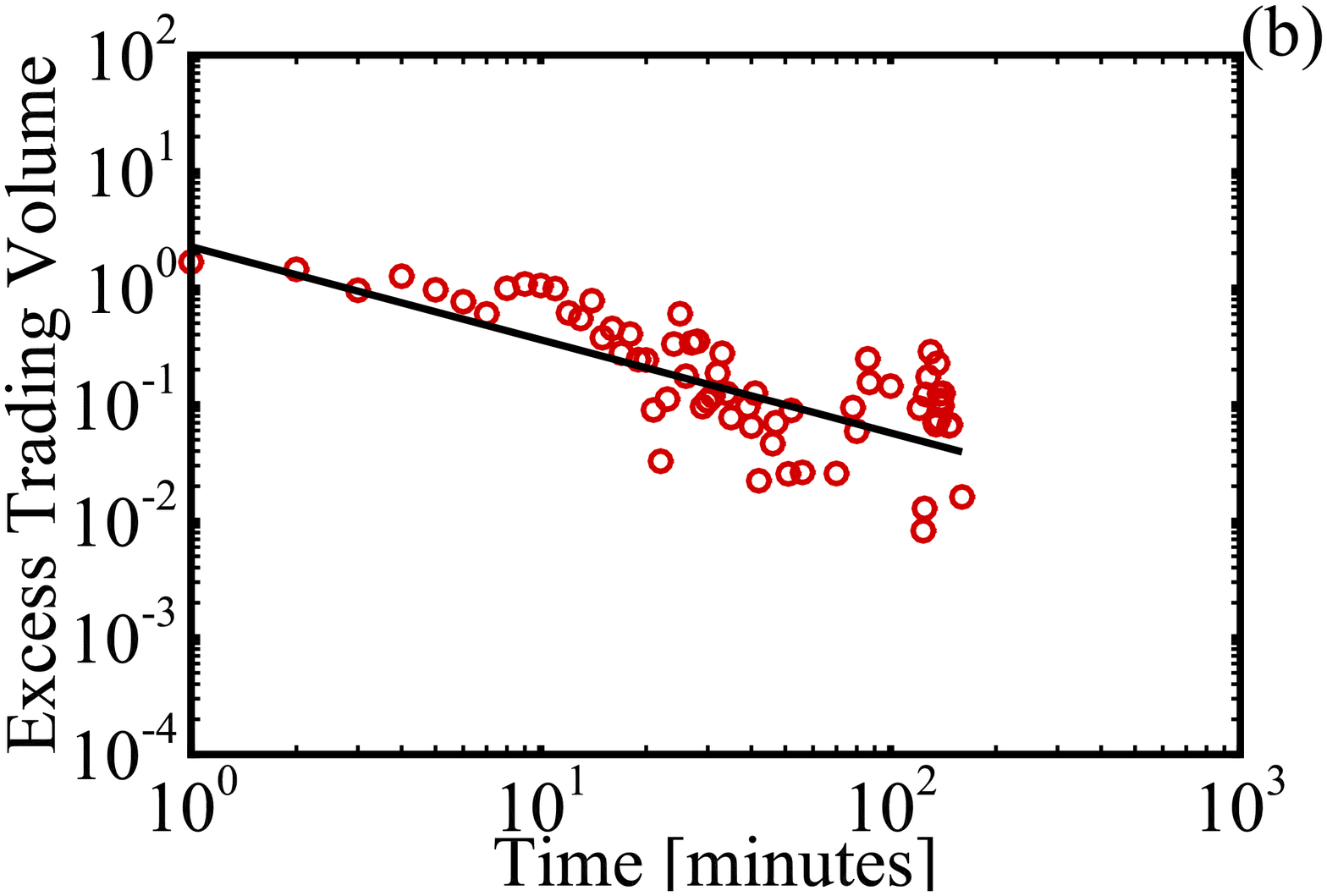}
  \includegraphics[width=0.32\textwidth]{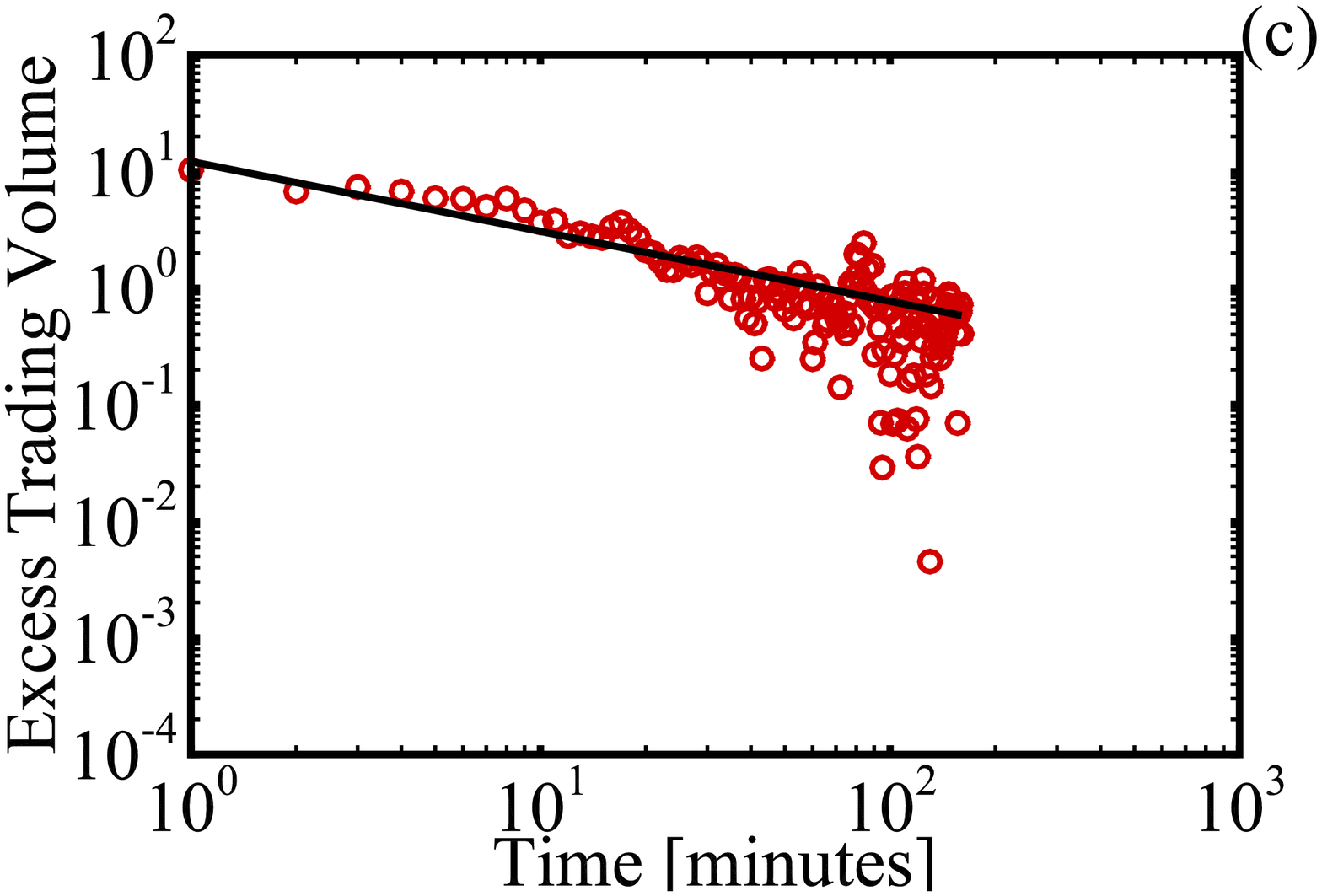}
  \includegraphics[width=0.32\textwidth]{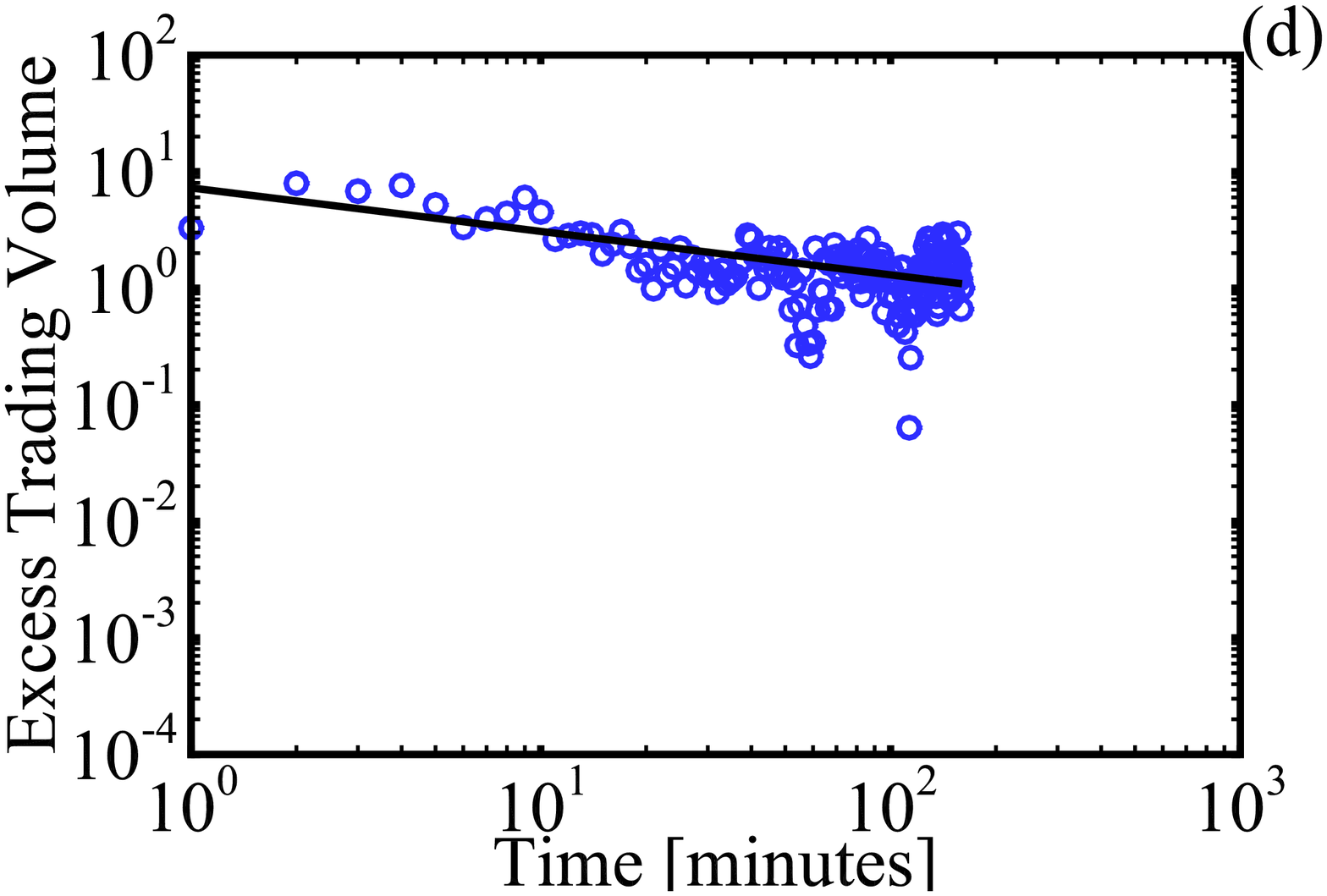}
  \includegraphics[width=0.32\textwidth]{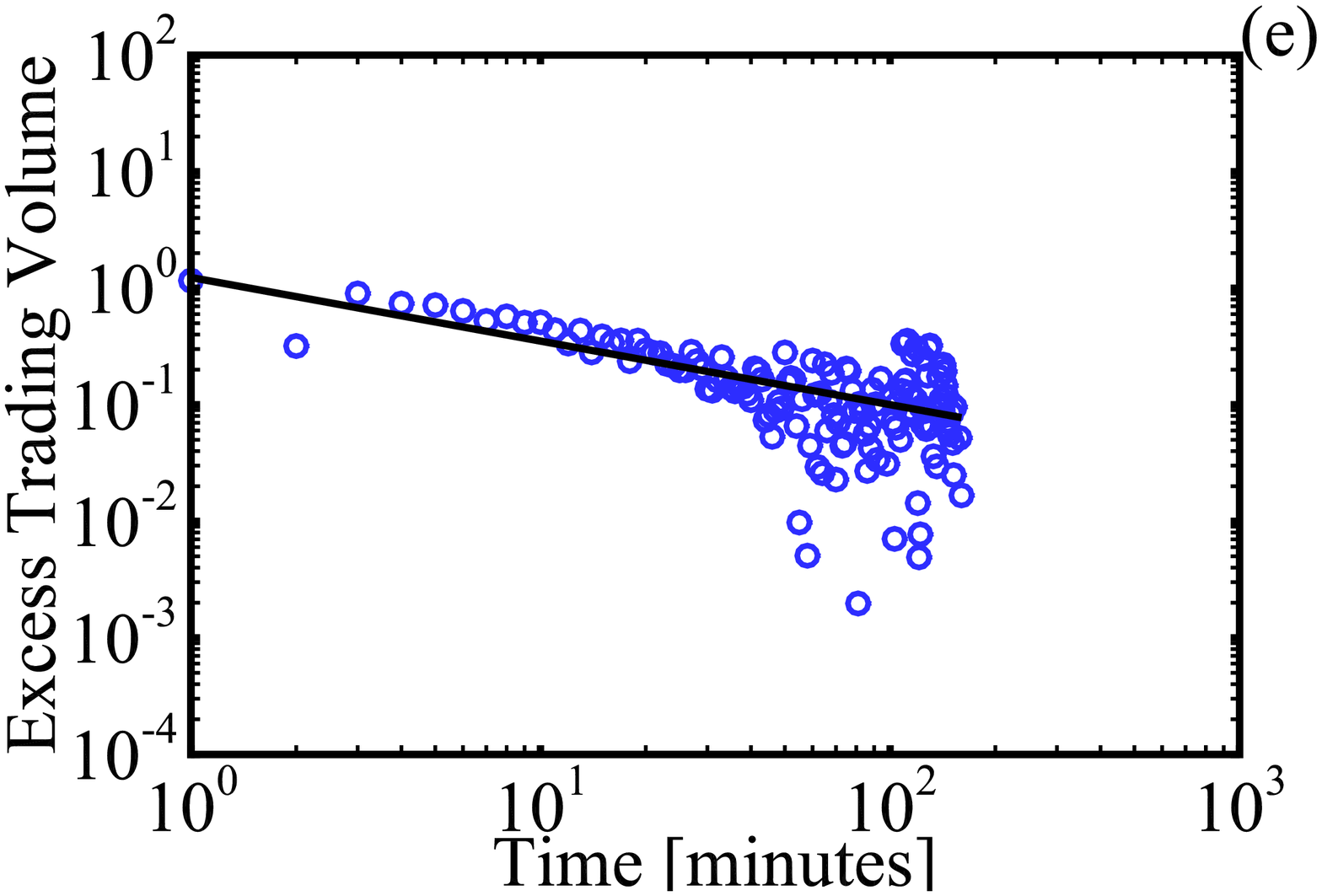}
  \includegraphics[width=0.32\textwidth]{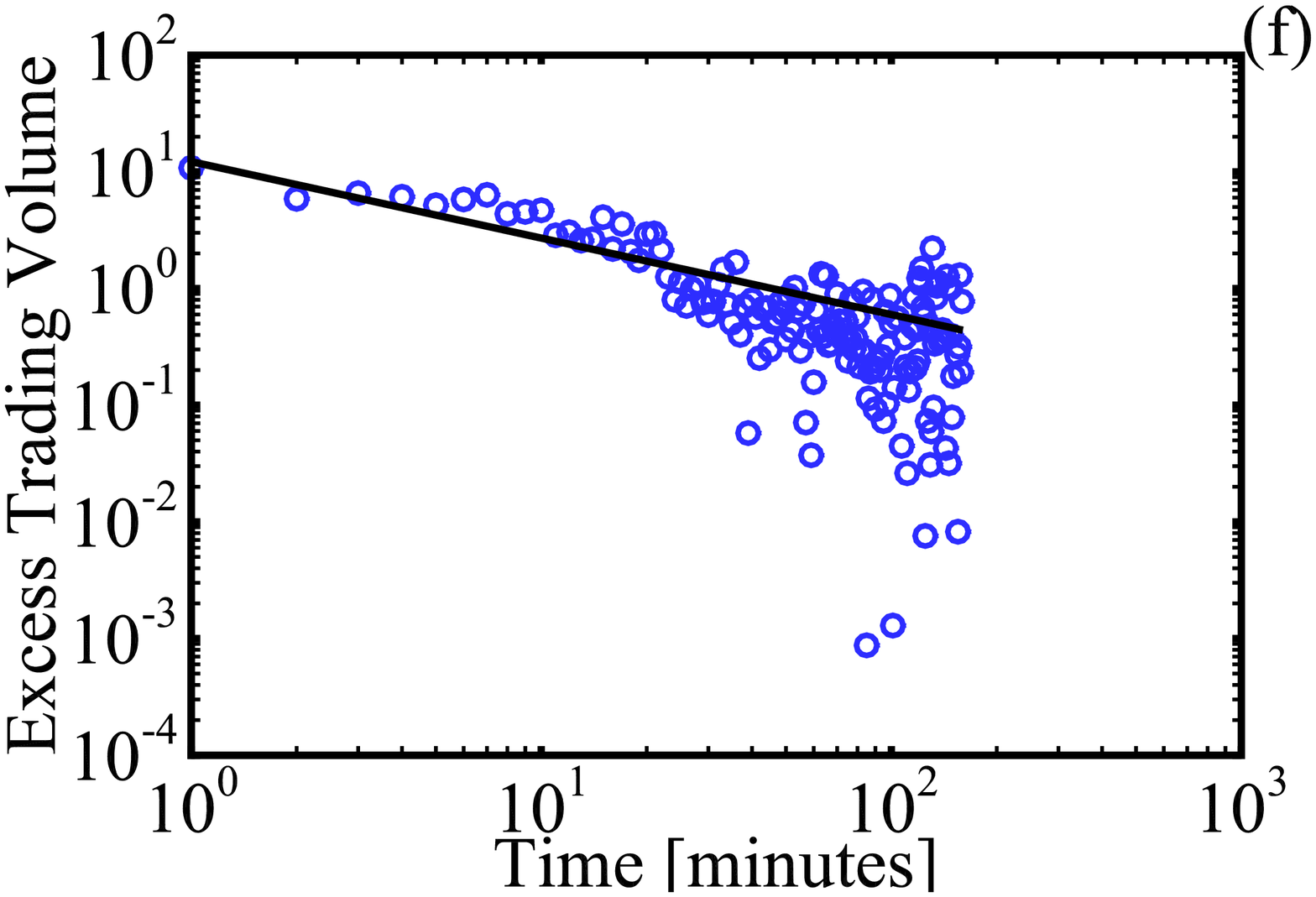}
  \caption{\label{Fig:PowerLaw:ExTraVol:Each} Power-law relaxation of excess variable for trading volume after three types of trading halts which are also divided into positive and negative events on log-log scales with power-law fits. The relaxation exponents $\alpha$ are: (a) 0.45 $\pm$ 0.03, (b) 0.80 $\pm$ 0.11, (c) 0.60 $\pm$ 0.03, (d) 0.37 $\pm$ 0.05, (e) 0.55 $\pm$ 0.05, and (f) 0.66 $\pm$ 0.04, respectively.}
\end{figure}

\begin{figure}[!htb]
  \centering
  \includegraphics[width=0.32\textwidth]{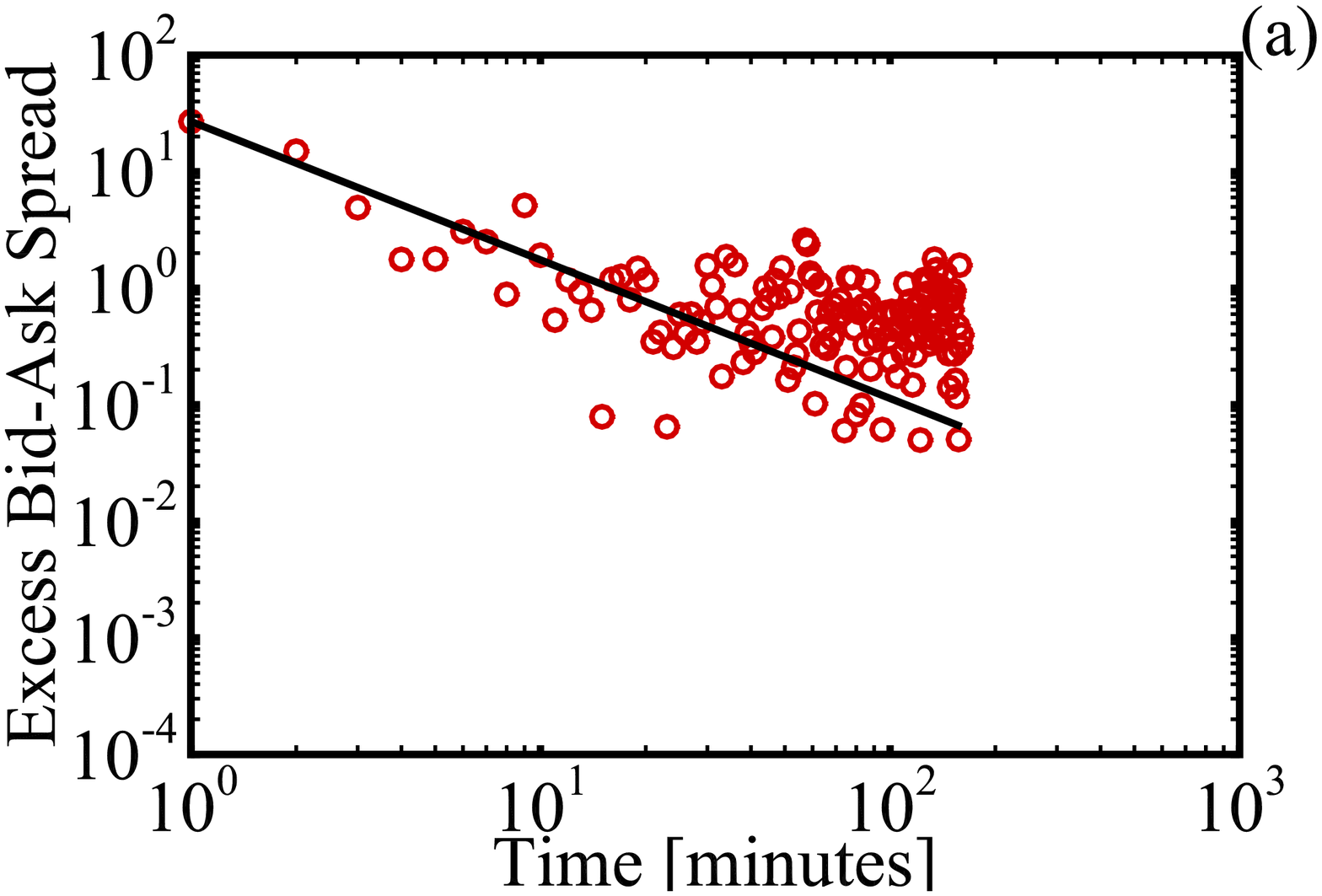}
  \includegraphics[width=0.32\textwidth]{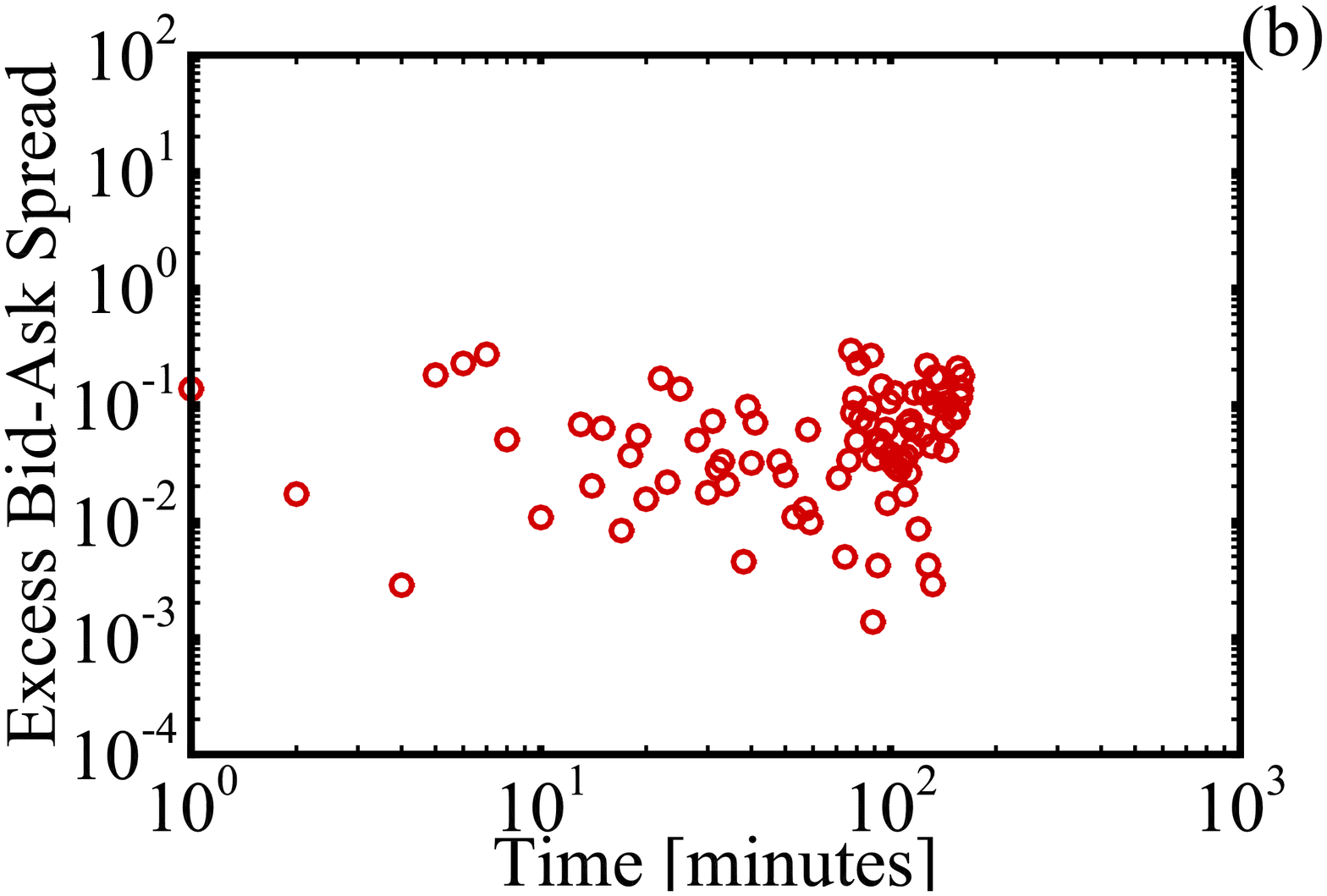}
  \includegraphics[width=0.32\textwidth]{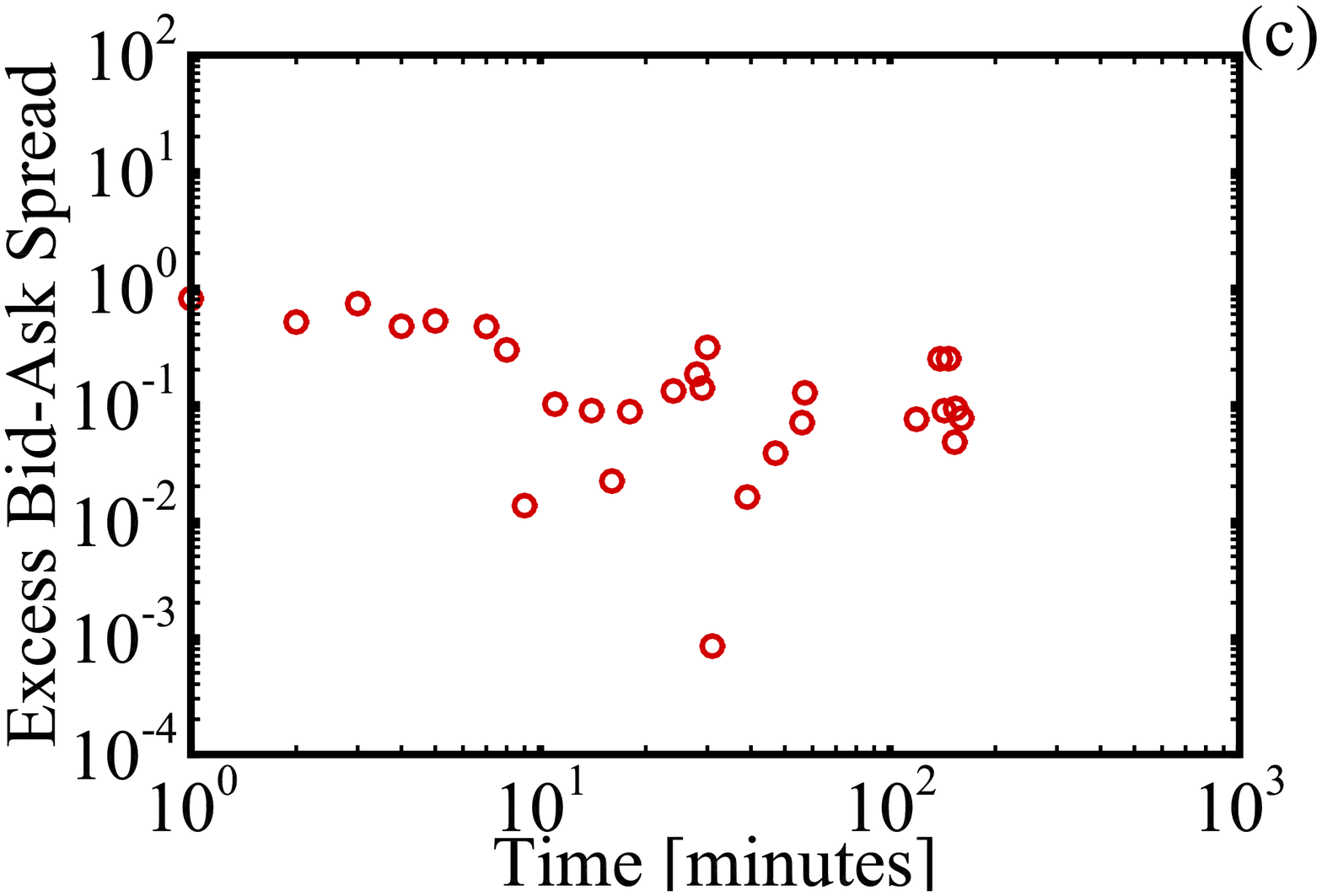}
  \includegraphics[width=0.32\textwidth]{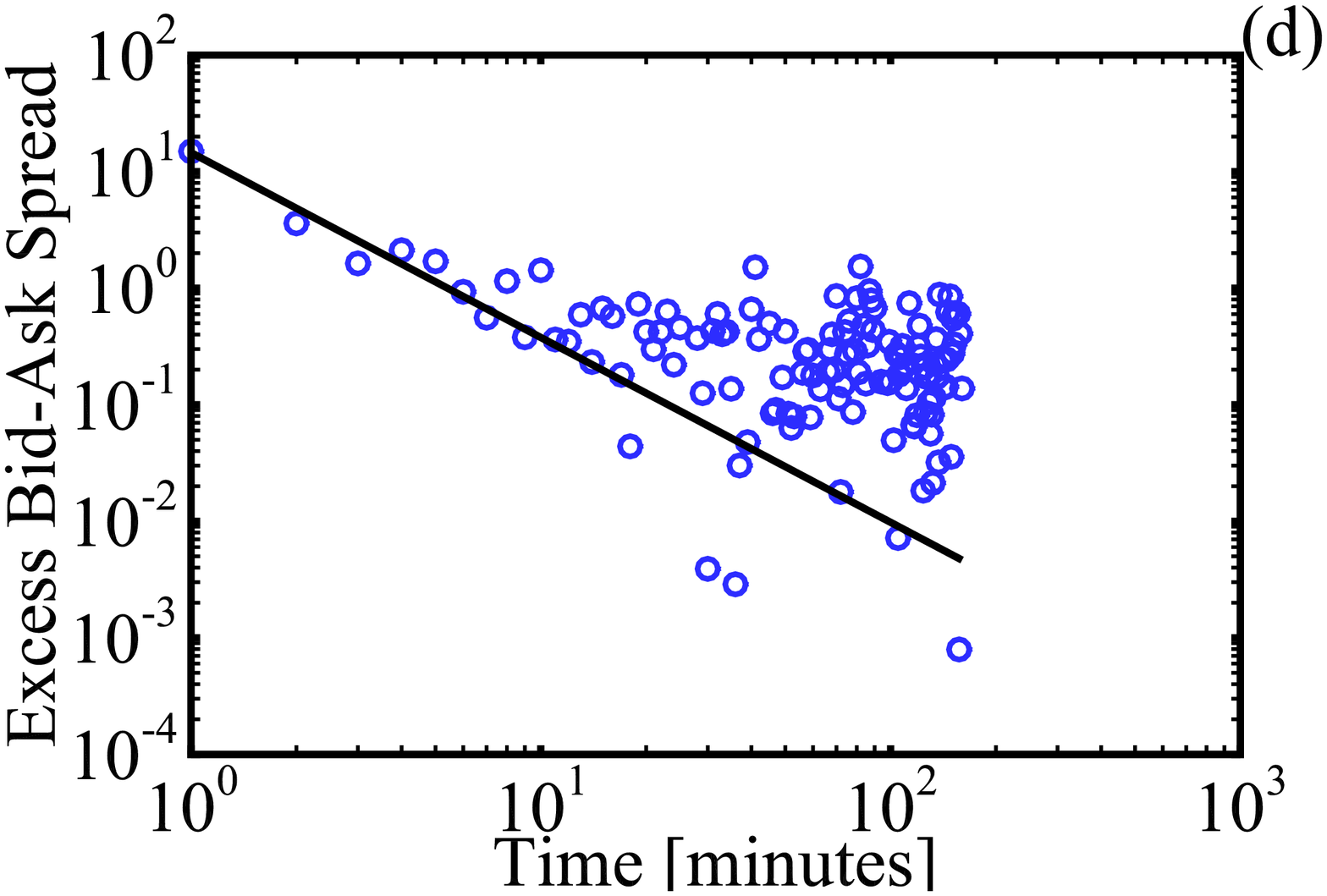}
  \includegraphics[width=0.32\textwidth]{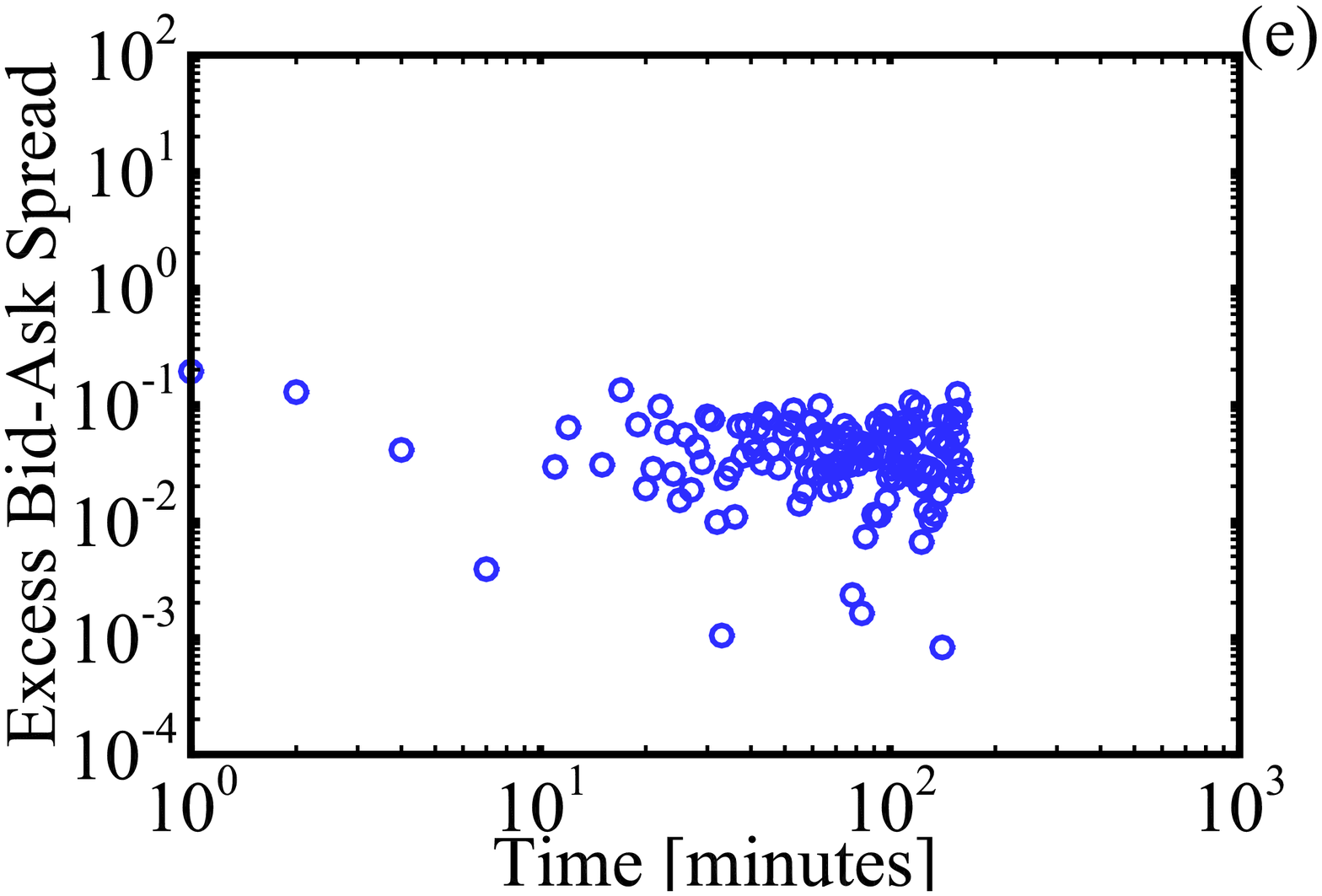}
  \includegraphics[width=0.32\textwidth]{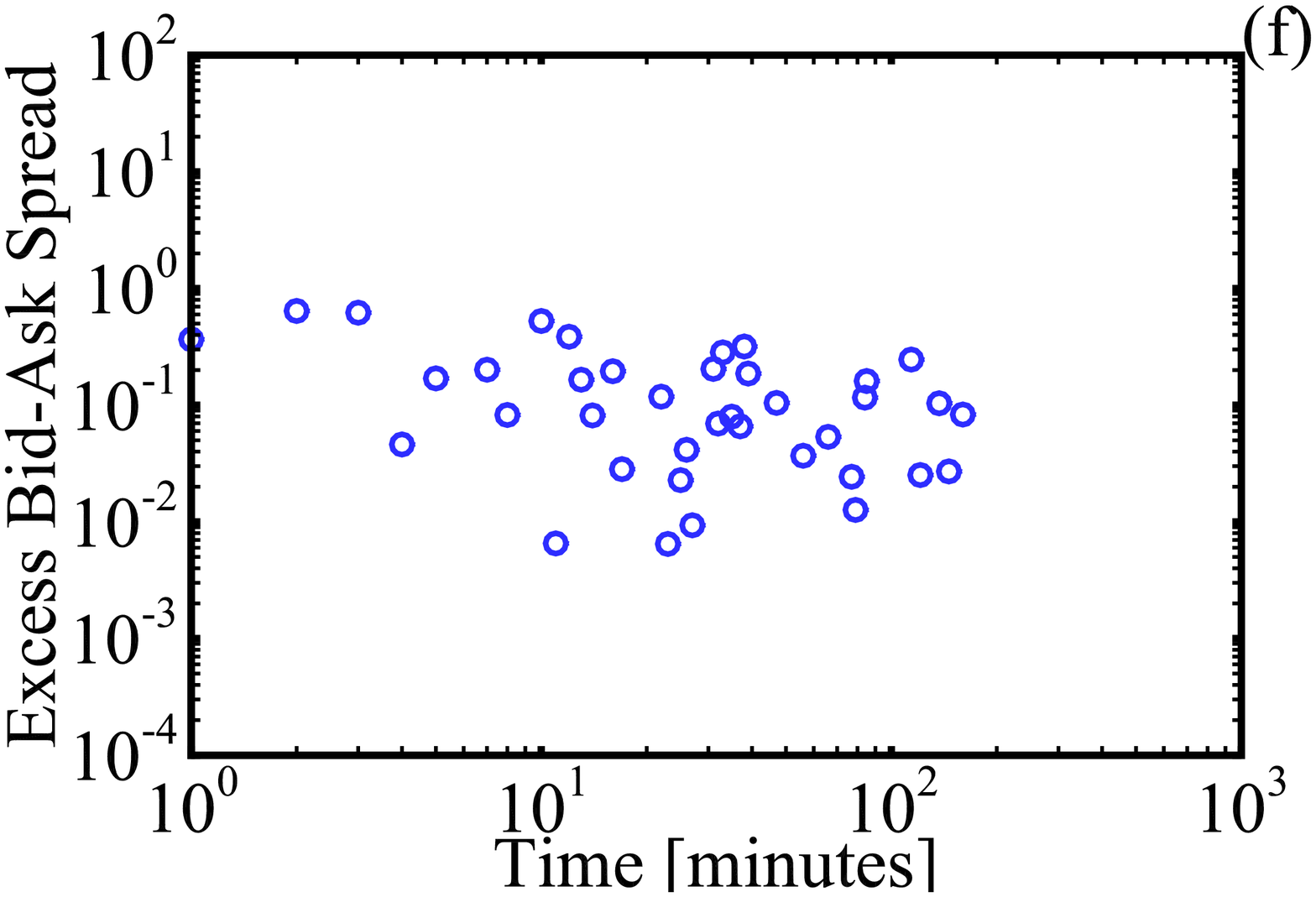}
  \caption{\label{Fig:PowerLaw:ExBidAskSpr:Each} Power-law relaxation of excess variable for bid-ask spread after three types of trading halts which are also divided into positive and negative events on log-log scales with power-law fits. The relaxation exponents $\alpha$ are: (a) 1.19 $\pm$ 0.08, and (d) 1.59 $\pm$ 0.13. There do not exhibit power-law relaxations in: (b) positive one-day halts, (c) positive inter-day halts, (e) negative one-day halts, and (f) negative inter-day halts.}
\end{figure}

To study relaxations of the three financial measures, we use the method of Ref. \cite{Mu-Zhou-Chen-Kertesz-2010-NJP} to plot the excess variables defined as the relative difference between the actual value and the corresponding value of intraday pattern $I(t)$,
\begin{equation}
 z_{ex}(t) = \frac{Z(t) - I(t)}{I(t)} = z(t) - 1,
 \label{Eq:ExcessVariables}
\end{equation}
which give the relative excess of the value in normal periods. Figures ~\ref{Fig:PowerLaw:ExAbsRet:Each}-~\ref{Fig:PowerLaw:ExBidAskSpr:Each} exhibit the relaxation of the excess variables after the trading halts on log-log scales corresponding to Figures~\ref{Fig:AbsRet:Each}-~\ref{Fig:BidAskSpr:Each} respectively. All the curves for excess absolute return, excess volume (and bid-ask spread in case of intraday halts) show power law behaviors,
\begin{equation}
 z_{ex}(t) \sim t^{-\alpha}, t > 0,
 \label{Eq:PowerLaw}
\end{equation}
which has the same form as the dynamics of other extreme events in Refs. \cite{Zawadowski-Kertesz-Andor-2004-PA,Zawadowski-Andor-Kertesz-2006-QF,Mu-Zhou-Chen-Kertesz-2010-NJP,Toth-Kertesz-Farmer-2009-EPJB}. The power law relaxation is more significant with less fluctuation and gentler slope for volume than for absolute return (and bid-ask spread in case of intraday halts).

The estimate of $\alpha$ can be obtained through the nonlinear least-squares regression and all the power law relaxation exponents are listed in Table~\ref{Tab:RelaxationExponents}. The relaxation exponents for excess volume are almost all less than those for excess absolute return, which indicates the trading volume decay slower than the absolute return as analyzed above. More importantly, in case of absolute return, the relaxation exponents of inter-day halts are larger than those of intraday halts and one-day halts for both positive events and negative events. This indicates that price changes after inter-day halts decay fastest and thus the inter-day halts show relatively more effective than intraday halts and one-day halts in the sense of relaxation speed. In addition, from the perspective of comparing positive events with negative events, in case of intraday halts and one-day halts the relaxation exponents of excess absolute return and excess volume for positive events are larger than those for negative events. In contrast, in case of inter-day halts, the relaxation exponents of excess absolute return and excess volume for positive events are smaller than those for negative events. This is partly due to higher peak appeared in positive events than in negative events for intraday halts and one-day halts, while opposite for inter-day halts.

\begin{table}[ht]
  \caption{\label{Tab:RelaxationExponents} Relaxation exponents of different types of trading halts.}
  \centering
  \begin{tabular}{c c c c c c c}
  \toprule
   & \multicolumn{2}{c}{Intraday halts} & \multicolumn{2}{c}{One-day halts} & \multicolumn{2}{c}{Inter-day halts}  \\ \cline{2-7}
  Variable        & Positive        & Negative        & Positive        & Negative        & Positive        & Negative        \\ \midrule
  Absolute return & 0.89 $\pm$ 0.05 & 0.61 $\pm$ 0.05 & 0.89 $\pm$ 0.13 & 0.54 $\pm$ 0.05 & 0.96 $\pm$ 0.07 & 1.15 $\pm$ 0.13 \\
  Volume          & 0.45 $\pm$ 0.03 & 0.37 $\pm$ 0.05 & 0.80 $\pm$ 0.11 & 0.55 $\pm$ 0.05 & 0.60 $\pm$ 0.03 & 0.66 $\pm$ 0.04 \\
  Bid-ask spread  & 1.19 $\pm$ 0.08 & 1.59 $\pm$ 0.13 & -               & -               & -               & -               \\
  \bottomrule
  \end{tabular}
\end{table}

\section{Conclusion}

In this article, we investigate the short-term market reaction after trading halts in Chinese stock market. By analyzing the dynamics of average cumulative return, we find trading halts prevent the sustained rising or falling and play a certain role in price discovery. While for different types of trading halts, the stabilities of cumulative return in after-halt period are different. The average cumulative returns after intraday halts are most volatile, while those after one-day halts are most stable. This difference may relate with the complexity of the information contained in the halt announcement.

By further analysis, we conclude that absolute return, trading volume, and in case of bid-ask spread around intraday halts all share the same pattern with a sharp peak and a power law relaxation after that, which is consistent with the typical signature of an exogenous shock \cite{Crane-Sornette-2008-PNAS}. While for different types of trading halts, the peaks' height and the relaxation exponents are different. From the perspective of price trends, the peak of positive events is higher than that of negative events for all these three measures. This can be explained by traders' asymmetric psychology. Furthermore, the relaxation exponents of excess absolute return and excess volume for positive events are larger than those for negative events in case of intraday halts and one-day halts. This means that positive events are more effective than negative events for intraday halts and one-day halts. In contrast, the relaxation exponents of excess absolute return and excess volume for positive events are smaller than those for negative events in case of inter-day halts. This means that negative events are more effective than positive events for inter-day halts. From the perspective of halt reasons or halt durations, intraday halts show the highest peak, and one-day halts show the lowest peak. In addition, the relaxation exponents of excess absolute return after inter-day halts are larger than those after intraday halts and one-day halts, which implies that inter-day halts are relatively more effective than intraday halts and one-day halts.

These financial dynamics give us a better understanding of the market reaction after different types of trading halts. Also, comparative analysis based on these patterns can help supervisors to easily assess the effects of different trading halts and improve suspension rules. Of course, there are more things worthy of further study. For example, now we can only speculate the origin of these differences between different types of trading halts. It may be related to behavioral trading or information content around the trading halts. This needs to be further conformed. More importantly, researchers or investors can employ these different patterns to develop different ``trading halts strategies''. Which type of halts is suitable for momentum strategies? Which type of halts is suitable for contrarian strategies? Are these strategies profitable? All these need to be further tested.

\bigskip
{\textbf{Acknowledgments:}}

The preliminary version of this paper was presented on the Econophysics Colloquium 2013 and Asia Pacific Econophysics Conference (APEC) 2013, Pohang, Korea. The authors thank Wei-Xing Zhou and other participants of this conference for helpful discussion. This work was supported by the National Natural Science Foundation of China under Key Project Grant No 71131007 and the Programme of Innovative Research Team supported by Ministry of Education of China under Grant No IRT1028.

\bigskip
{\textbf{Appendix:}}

 The data sets are composed by 1-min high-frequency trading data of 203 stocks traded on Shanghai Stock Exchange (SSE). These stocks are selected based on scale and liquidity and cover a variety of industry sectors. The tickers of these 203 stocks are as following: 600000, 600005, 600008, 600009, 600010, 600015, 600016, 600019, 600026, 600028, 600029, 600030, 600031, 600036, 600037, 600048, 600050, 600058, 600062, 600066, 600068, 600085, 600089, 600096, 600098, 600100, 600104, 600108, 600109, 600111, 600115, 600118, 600123, 600125, 600132, 600143, 600150, 600151, 600153, 600160, 600161, 600166, 600169, 600170, 600177, 600183, 600188, 600196, 600208, 600216, 600219, 600221, 600252, 600256, 600259, 600266, 600267, 600271, 600276, 600307, 600309, 600316, 600320, 600331, 600348, 600352, 600362, 600369, 600372, 600376, 600380, 600383, 600395, 600406, 600415, 600418, 600428, 600432, 600456, 600481, 600489, 600497, 600498, 600500, 600508, 600516, 600518, 600519, 600528, 600535, 600546, 600547, 600549, 600550, 600582, 600583, 600585, 600588, 600595, 600598, 600600, 600635, 600642, 600649, 600655, 600664, 600674, 600690, 600694, 600703, 600718, 600737, 600739, 600741, 600770, 600779, 600783, 600795, 600804, 600808, 600809, 600811, 600812, 600832, 600837, 600839, 600859, 600863, 600873, 600875, 600879, 600881, 600887, 600893, 600895, 600900, 600970, 600971, 600997, 600999, 601001, 601006, 601009, 601018, 601088, 601098, 601099, 601101, 601106, 601111, 601117, 601118, 601158, 601166, 601168, 601169, 601179, 601186, 601216, 601233, 601268, 601288, 601299, 601318, 601328, 601333, 601369, 601377, 601390, 601398, 601519, 601558, 601600, 601601, 601607, 601618, 601628, 601666, 601668, 601688, 601699, 601717, 601718, 601727, 601766, 601788, 601808, 601818, 601857, 601866, 601888, 601898, 601899, 601918, 601919, 601933, 601939, 601958, 601988, 601989, 601991, 601992, and 601998.

\bibliography{Short_v2}

\begin{thebibliography}{33}
\expandafter\ifx\csname natexlab\endcsname\relax\def\natexlab#1{#1}\fi
\providecommand{\url}[1]{\texttt{#1}}
\providecommand{\href}[2]{#2}
\providecommand{\path}[1]{#1}
\providecommand{\DOIprefix}{doi:}
\providecommand{\ArXivprefix}{arXiv:}
\providecommand{\URLprefix}{URL: }
\providecommand{\Pubmedprefix}{pmid:}
\providecommand{\doi}[1]{\href{http://dx.doi.org/#1}{\path{#1}}}
\providecommand{\Pubmed}[1]{\href{pmid:#1}{\path{#1}}}
\providecommand{\bibinfo}[2]{#2}
\ifx\xfnm\relax \def\xfnm[#1]{\unskip,\space#1}\fi
\bibitem[{Sornette(2009)}]{Sornette-2009-PUP}
\bibinfo{author}{D.~Sornette}, \bibinfo{title}{Why stock markets crash:
  critical events in complex financial systems}, \bibinfo{publisher}{Princeton
  University Press}, \bibinfo{year}{2009}.
\bibitem[{Lillo and Mantegna(2003)}]{Lillo-Mantegna-2003-PRE}
\bibinfo{author}{F.~Lillo}, \bibinfo{author}{R.~N. Mantegna},
\newblock \bibinfo{title}{Power-law relaxation in a complex system: Omori law
  after a financial market crash},
\newblock \bibinfo{journal}{Phys. Rev. E} \bibinfo{volume}{68}
  (\bibinfo{year}{2003}) \bibinfo{pages}{016119}.
\bibitem[{Lillo and Mantegna(2004)}]{Lillo-Mantegna-2004-PA}
\bibinfo{author}{F.~Lillo}, \bibinfo{author}{R.~N. Mantegna},
\newblock \bibinfo{title}{Dynamics of a financial market index after a crash},
\newblock \bibinfo{journal}{Physica A} \bibinfo{volume}{338}
  (\bibinfo{year}{2004}) \bibinfo{pages}{125--134}.
\bibitem[{Weber et~al.(2007)Weber, Wang, Vodenska-Chitkushev, Havlin, and
  Stanley}]{Weber-Wang-Vodenska-Havlin-Stanley-2007-PRE}
\bibinfo{author}{P.~Weber}, \bibinfo{author}{F.~Wang},
  \bibinfo{author}{I.~Vodenska-Chitkushev}, \bibinfo{author}{S.~Havlin},
  \bibinfo{author}{H.~E. Stanley},
\newblock \bibinfo{title}{Relation between volatility correlations in financial
  markets and {O}mori processes occurring on all scales},
\newblock \bibinfo{journal}{Phys. Rev. E} \bibinfo{volume}{76}
  (\bibinfo{year}{2007}) \bibinfo{pages}{016109}.
\bibitem[{Mu and Zhou(2008)}]{Mu-Zhou-2008-PA}
\bibinfo{author}{G.-H. Mu}, \bibinfo{author}{W.-X. Zhou},
\newblock \bibinfo{title}{Relaxation dynamics of aftershocks after large
  volatility shocks in the {SSEC} index},
\newblock \bibinfo{journal}{Physica A} \bibinfo{volume}{387}
  (\bibinfo{year}{2008}) \bibinfo{pages}{5211--5218}.
\bibitem[{Petersen et~al.(2010{\natexlab{a}})Petersen, Wang, Havlin, and
  Stanley}]{Petersen-Wang-Havlin-Stanley-2010-PRE-036114}
\bibinfo{author}{A.~M. Petersen}, \bibinfo{author}{F.~Wang},
  \bibinfo{author}{S.~Havlin}, \bibinfo{author}{H.~E. Stanley},
\newblock \bibinfo{title}{Market dynamics immediately before and after
  financial shocks: Quantifying the {O}mori, productivity, and {B}ath laws},
\newblock \bibinfo{journal}{Phys. Rev. E} \bibinfo{volume}{82}
  (\bibinfo{year}{2010}{\natexlab{a}}) \bibinfo{pages}{036114}.
\bibitem[{Petersen et~al.(2010{\natexlab{b}})Petersen, Wang, Havlin, and
  Stanley}]{Petersen-Wang-Havlin-Stanley-2010-PRE-066121}
\bibinfo{author}{A.~M. Petersen}, \bibinfo{author}{F.~Wang},
  \bibinfo{author}{S.~Havlin}, \bibinfo{author}{H.~E. Stanley},
\newblock \bibinfo{title}{Quantitative law describing market dynamics before
  and after interest-rate change},
\newblock \bibinfo{journal}{Phys. Rev. E} \bibinfo{volume}{81}
  (\bibinfo{year}{2010}{\natexlab{b}}) \bibinfo{pages}{066121}.
\bibitem[{Zawadowski et~al.(2004)Zawadowski, Kert\'esz, and
  Andor}]{Zawadowski-Kertesz-Andor-2004-PA}
\bibinfo{author}{A.~Zawadowski}, \bibinfo{author}{J.~Kert\'esz},
  \bibinfo{author}{G.~Andor},
\newblock \bibinfo{title}{Large price changes on small scales},
\newblock \bibinfo{journal}{Physica A} \bibinfo{volume}{344}
  (\bibinfo{year}{2004}) \bibinfo{pages}{221--226}.
\bibitem[{Zawadowski et~al.(2006)Zawadowski, Andor, and
  Kert\'esz}]{Zawadowski-Andor-Kertesz-2006-QF}
\bibinfo{author}{A.~G. Zawadowski}, \bibinfo{author}{G.~Andor},
  \bibinfo{author}{J.~Kert\'esz},
\newblock \bibinfo{title}{Short-term market reaction after extreme price
  changes of liquid stocks},
\newblock \bibinfo{journal}{Quant. Finance} \bibinfo{volume}{6}
  (\bibinfo{year}{2006}) \bibinfo{pages}{283--295}.
\bibitem[{Ponzi et~al.(2009)Ponzi, Lillo, and
  Mantegna}]{Ponzi-Lillo-Mantegna-2009-PRE}
\bibinfo{author}{A.~Ponzi}, \bibinfo{author}{F.~Lillo}, \bibinfo{author}{R.~N.
  Mantegna},
\newblock \bibinfo{title}{Market reaction to a bid-ask spread change: A
  power-law relaxation dynamics},
\newblock \bibinfo{journal}{Phys. Rev. E} \bibinfo{volume}{80}
  (\bibinfo{year}{2009}) \bibinfo{pages}{016112}.
\bibitem[{Mu et~al.(2010)Mu, Zhou, Chen, and
  Kert\'esz}]{Mu-Zhou-Chen-Kertesz-2010-NJP}
\bibinfo{author}{G.~H. Mu}, \bibinfo{author}{W.~X. Zhou},
  \bibinfo{author}{W.~Chen}, \bibinfo{author}{J.~Kert\'esz},
\newblock \bibinfo{title}{Order flow dynamics around extreme price changes on
  an emerging stock market},
\newblock \bibinfo{journal}{New J. Phys.} \bibinfo{volume}{12}
  (\bibinfo{year}{2010}) \bibinfo{pages}{075037}.
\bibitem[{T\'oth et~al.(2009)T\'oth, Kert\'esz, and
  Farmer}]{Toth-Kertesz-Farmer-2009-EPJB}
\bibinfo{author}{B.~T\'oth}, \bibinfo{author}{J.~Kert\'esz},
  \bibinfo{author}{J.~D. Farmer},
\newblock \bibinfo{title}{Studies of the limit order book around large price
  changes},
\newblock \bibinfo{journal}{Eur. Phys. J. B} \bibinfo{volume}{71}
  (\bibinfo{year}{2009}) \bibinfo{pages}{499--510}.
\bibitem[{Stanley et~al.(2010)Stanley, Buldyrev, Franzese, Havlin, Mallamace,
  Kumar, Plerou, and Preis}]{Stanley-Buldyrev-Franzese-2010-PA}
\bibinfo{author}{H.~Stanley}, \bibinfo{author}{S.~Buldyrev},
  \bibinfo{author}{G.~Franzese}, \bibinfo{author}{S.~Havlin},
  \bibinfo{author}{F.~Mallamace}, \bibinfo{author}{P.~Kumar},
  \bibinfo{author}{V.~Plerou}, \bibinfo{author}{T.~Preis},
\newblock \bibinfo{title}{Correlated randomness and switching phenomena},
\newblock \bibinfo{journal}{Physica A} \bibinfo{volume}{389}
  (\bibinfo{year}{2010}) \bibinfo{pages}{2880--2893}.
\bibitem[{Preis and Stanley(2010)}]{Preis-Stanley-2010-JSP}
\bibinfo{author}{T.~Preis}, \bibinfo{author}{H.~E. Stanley},
\newblock \bibinfo{title}{Switching phenomena in a system with no switches},
\newblock \bibinfo{journal}{J. Stat. Phys.} \bibinfo{volume}{138}
  (\bibinfo{year}{2010}) \bibinfo{pages}{431--446}.
\bibitem[{Preis et~al.(2011)Preis, Schneider, and
  Stanley}]{Preis-Schneider-Stanley-2011-PNAS}
\bibinfo{author}{T.~Preis}, \bibinfo{author}{J.~J. Schneider},
  \bibinfo{author}{H.~E. Stanley},
\newblock \bibinfo{title}{Switching processes in financial markets},
\newblock \bibinfo{journal}{Proc. Natl. Acad. Sci. USA} \bibinfo{volume}{108}
  (\bibinfo{year}{2011}) \bibinfo{pages}{7674--7678}.
\bibitem[{Jiang et~al.(2013)Jiang, Chen, and Zheng}]{Jiang-Chen-Zheng-2013-PA}
\bibinfo{author}{X.~Jiang}, \bibinfo{author}{T.~Chen},
  \bibinfo{author}{B.~Zheng},
\newblock \bibinfo{title}{Time-reversal asymmetry in financial systems},
\newblock \bibinfo{journal}{Physica A} \bibinfo{volume}{392}
  (\bibinfo{year}{2013}) \bibinfo{pages}{5369--5375}.
\bibitem[{Sornette and Helmstetter(2003)}]{Sornette-Helmstetter-2003-PA}
\bibinfo{author}{D.~Sornette}, \bibinfo{author}{A.~Helmstetter},
\newblock \bibinfo{title}{Endogenous versus exogenous shocks in systems with
  memory},
\newblock \bibinfo{journal}{Physica A} \bibinfo{volume}{318}
  (\bibinfo{year}{2003}) \bibinfo{pages}{577--591}.
\bibitem[{Sornette et~al.(2004)Sornette, Deschatres, Gilbert, and
  Ageon}]{Sornette-Deschatres-Gilbert-Ageon-2004-PRL}
\bibinfo{author}{D.~Sornette}, \bibinfo{author}{F.~Deschatres},
  \bibinfo{author}{T.~Gilbert}, \bibinfo{author}{Y.~Ageon},
\newblock \bibinfo{title}{Endogenous versus exogenous shocks in complex
  networks: An empirical test using book sale rankings},
\newblock \bibinfo{journal}{Phys. Rev. Lett.} \bibinfo{volume}{93}
  (\bibinfo{year}{2004}) \bibinfo{pages}{228701}.
\bibitem[{Roehner et~al.(2004)Roehner, Sornette, and
  Andersen}]{Roehner-Sornette-Andersen-2004-IJMPC}
\bibinfo{author}{B.~Roehner}, \bibinfo{author}{D.~Sornette},
  \bibinfo{author}{J.~Andersen},
\newblock \bibinfo{title}{Response functions to critical shocks in social
  sciences: An empirical and numerical study},
\newblock \bibinfo{journal}{Int. J. Mod. Phys. C} \bibinfo{volume}{15}
  (\bibinfo{year}{2004}) \bibinfo{pages}{809--834}.
\bibitem[{Deschatres and Sornette(2005)}]{Deschatres-Sornette-2005-PRE}
\bibinfo{author}{F.~Deschatres}, \bibinfo{author}{D.~Sornette},
\newblock \bibinfo{title}{Dynamics of book sales: Endogenous versus exogenous
  shocks in complex networks},
\newblock \bibinfo{journal}{Phys. Rev. E} \bibinfo{volume}{72}
  (\bibinfo{year}{2005}) \bibinfo{pages}{016112}.
\bibitem[{Crane and Sornette(2008)}]{Crane-Sornette-2008-PNAS}
\bibinfo{author}{R.~Crane}, \bibinfo{author}{D.~Sornette},
\newblock \bibinfo{title}{Robust dynamic classes revealed by measuring the
  response function of a social system},
\newblock \bibinfo{journal}{Proc. Natl. Acad. Sci. USA} \bibinfo{volume}{105}
  (\bibinfo{year}{2008}) \bibinfo{pages}{15649--15653}.
\bibitem[{Lee et~al.(1994)Lee, Ready, and Seguin}]{Lee-Ready-Seguin-1994-JF}
\bibinfo{author}{C.~Lee}, \bibinfo{author}{M.~J. Ready}, \bibinfo{author}{P.~J.
  Seguin},
\newblock \bibinfo{title}{Volume, volatility, and {N}ew {Y}ork stock exchange
  trading halts},
\newblock \bibinfo{journal}{J. Finance} \bibinfo{volume}{49}
  (\bibinfo{year}{1994}) \bibinfo{pages}{183--214}.
\bibitem[{Subrahmanyam(1994)}]{Subrahmanyam-1994-JF}
\bibinfo{author}{A.~Subrahmanyam},
\newblock \bibinfo{title}{Circuit breakers and market volatility: A theoretical
  perspective},
\newblock \bibinfo{journal}{J. Finance} \bibinfo{volume}{49}
  (\bibinfo{year}{1994}) \bibinfo{pages}{237--254}.
\bibitem[{Corwin and Lipson(2000)}]{Corwin-Lipson-2000-JF}
\bibinfo{author}{S.~A. Corwin}, \bibinfo{author}{M.~L. Lipson},
\newblock \bibinfo{title}{Order flow and liquidity around {NYSE} trading
  halts},
\newblock \bibinfo{journal}{J. Finance} \bibinfo{volume}{55}
  (\bibinfo{year}{2000}) \bibinfo{pages}{1771--1805}.
\bibitem[{Ackert et~al.(2001)Ackert, Church, and
  Jayaraman}]{Ackert-Church-Jayaraman-2001-JFM}
\bibinfo{author}{L.~F. Ackert}, \bibinfo{author}{B.~Church},
  \bibinfo{author}{N.~Jayaraman},
\newblock \bibinfo{title}{An experimental study of circuit breakers: the
  effects of mandated market closures and temporary halts on market behavior},
\newblock \bibinfo{journal}{J. Financ. Mark} \bibinfo{volume}{4}
  (\bibinfo{year}{2001}) \bibinfo{pages}{185--208}.
\bibitem[{Christie et~al.(2002)Christie, Corwin, and
  Harris}]{Christie-Corwin-Harris-2002-JF}
\bibinfo{author}{W.~G. Christie}, \bibinfo{author}{S.~A. Corwin},
  \bibinfo{author}{J.~H. Harris},
\newblock \bibinfo{title}{Nasdaq trading halts: The impact of market mechanisms
  on prices, trading activity, and execution costs},
\newblock \bibinfo{journal}{J. Finance} \bibinfo{volume}{57}
  (\bibinfo{year}{2002}) \bibinfo{pages}{1443--1478}.
\bibitem[{Edelen and Gervais(2003)}]{Edelen-Gervais-2003-RFS}
\bibinfo{author}{R.~Edelen}, \bibinfo{author}{S.~Gervais},
\newblock \bibinfo{title}{The role of trading halts in monitoring a specialist
  market},
\newblock \bibinfo{journal}{Rev. Finan. Stud.} \bibinfo{volume}{16}
  (\bibinfo{year}{2003}) \bibinfo{pages}{263--300}.
\bibitem[{Engelen and Kabir(2006)}]{Engelen-Kabir-2006-JBFA}
\bibinfo{author}{P.~J. Engelen}, \bibinfo{author}{R.~Kabir},
\newblock \bibinfo{title}{Empirical evidence on the role of trading suspensions
  in disseminating new information to the capital market},
\newblock \bibinfo{journal}{J. Bus. Finan. Account} \bibinfo{volume}{33}
  (\bibinfo{year}{2006}) \bibinfo{pages}{1142--1167}.
\bibitem[{Hauser et~al.(2006)Hauser, Kedar-Levy, Pilo, and
  Shurki}]{Hauser-Kedar-Pilo-Shurki-2006-JFSR}
\bibinfo{author}{S.~Hauser}, \bibinfo{author}{H.~Kedar-Levy},
  \bibinfo{author}{B.~Pilo}, \bibinfo{author}{I.~Shurki},
\newblock \bibinfo{title}{The effect of trading halts on the speed of price
  discovery},
\newblock \bibinfo{journal}{J. Finan. Services Res.} \bibinfo{volume}{29}
  (\bibinfo{year}{2006}) \bibinfo{pages}{83--99}.
\bibitem[{Madura et~al.(2006)Madura, Richie, and
  Tucker}]{Madura-Richie-Tucker-2006-JFSR}
\bibinfo{author}{J.~Madura}, \bibinfo{author}{N.~Richie},
  \bibinfo{author}{A.~L. Tucker},
\newblock \bibinfo{title}{Trading halts and price discovery},
\newblock \bibinfo{journal}{J. Finan. Services Res.} \bibinfo{volume}{30}
  (\bibinfo{year}{2006}) \bibinfo{pages}{311--328}.
\bibitem[{Jiang et~al.(2009)Jiang, McInish, and
  Upson}]{Jiang-McInish-Upson-2009-JFM}
\bibinfo{author}{C.~Jiang}, \bibinfo{author}{T.~McInish},
  \bibinfo{author}{J.~Upson},
\newblock \bibinfo{title}{The information content of trading halts},
\newblock \bibinfo{journal}{J. Financ. Mark} \bibinfo{volume}{12}
  (\bibinfo{year}{2009}) \bibinfo{pages}{703--726}.
\bibitem[{Chakrabarty et~al.(2011)Chakrabarty, Corwin, and
  Panayides}]{Chakrabarty-Corwin-Panayides-2011-JFI}
\bibinfo{author}{B.~Chakrabarty}, \bibinfo{author}{S.~A. Corwin},
  \bibinfo{author}{M.~A. Panayides},
\newblock \bibinfo{title}{When a halt is not a halt: An analysis of off-{NYSE}
  trading during {NYSE} market closures},
\newblock \bibinfo{journal}{J. Finan. Intermediation} \bibinfo{volume}{20}
  (\bibinfo{year}{2011}) \bibinfo{pages}{361--386}.
\bibitem[{Kim and Yang(2004)}]{kim2004makes}
\bibinfo{author}{Y.~H. Kim}, \bibinfo{author}{J.~J. Yang},
\newblock \bibinfo{title}{What makes circuit breakers attractive to financial
  markets? a survey},
\newblock \bibinfo{journal}{Finan. Markets, Inst. Instruments}
  \bibinfo{volume}{13} (\bibinfo{year}{2004}) \bibinfo{pages}{109--146}.

\end{thebibliography}

\end{document}